\documentclass[10pt,twocolumn,twoside,letterpaper]{IEEEtran}
\makeatletter
\long\def\@makecaption#1#2{\ifx\@captype\@IEEEtablestring%
	\footnotesize\begin{center}{\normalfont\footnotesize #1}\\
		{\normalfont\footnotesize\scshape #2}\end{center}%
	\@IEEEtablecaptionsepspace
	\else
	\@IEEEfigurecaptionsepspace
	\setbox\@tempboxa\hbox{\normalfont\footnotesize {#1.}~~ #2}%
	\ifdim \wd\@tempboxa >\hsize%
	\setbox\@tempboxa\hbox{\normalfont\footnotesize {#1.}~~ }%
	\parbox[t]{\hsize}{\normalfont\footnotesize \noindent\unhbox\@tempboxa#2}%
	\else
	\hbox to\hsize{\normalfont\footnotesize\hfil\box\@tempboxa\hfil}\fi\fi}
\makeatother

\usepackage{cite}
\usepackage[dvips]{graphicx}

\usepackage[cmex10]{amsmath}
\usepackage{amssymb}
\usepackage[ruled,linesnumbered]{algorithm2e}
\usepackage{array}
\usepackage{mdwmath}
\usepackage{mdwtab}
\usepackage{eqparbox}
\usepackage{multirow}
\usepackage[tight,footnotesize]{subfigure}
\usepackage{stfloats}
\usepackage{url}
\usepackage{rotating,setspace}

\begin{document}

\title{Robust ENF Estimation Based on Harmonic Enhancement and Maximum Weight Clique}

\author{Guang Hua, \IEEEmembership{Member, IEEE}, Han Liao, Haijian Zhang, \IEEEmembership{Member, IEEE},\\ Dengpan Ye, \IEEEmembership{Member, IEEE}, and Jiayi Ma, \IEEEmembership{Member, IEEE}
	\thanks{G. Hua, H. Liao, H. Zhang, and J. Ma are with the School of Electronic Information, Wuhan University, Wuhan 430072, China (e-mail: \{ghua, liaohan, haijian.zhang\}@whu.edu.cn; jyma2010@gmail.com).}
	\thanks{D. Ye is with the School of Cyber Science and Engineering, Wuhan University, Wuhan 430072, China  (e-mail: yedp@whu.edu.cn).}
}

\maketitle

\begin{abstract}
We present a framework for robust electric network frequency (ENF) extraction from real-world audio recordings, featuring multi-tone ENF harmonic enhancement and graph-based optimal harmonic selection. Specifically, We first extend the recently developed single-tone ENF signal enhancement method to the multi-tone scenario and propose a harmonic robust filtering algorithm (HRFA). It can respectively enhance each harmonic component without cross-component interference, thus further alleviating the effects of unwanted noise and audio content on the much weaker ENF signal. In addition, considering the fact that some harmonic components could be severely corrupted even after enhancement, disturbing rather than facilitating ENF estimation, we propose a graph-based harmonic selection algorithm (GHSA), which finds the optimal combination of harmonic components for more accurate ENF estimation. Noticeably, the harmonic selection problem is equivalently formulated as a maximum weight clique (MWC) problem in graph theory, and the Bron-Kerbosch algorithm (BKA) is adopted in the GHSA. With the enhanced and optimally selected harmonic components, both the existing maximum likelihood estimator (MLE)  and weighted MLE (WMLE) are incorporated to yield the final ENF estimation results. The proposed framework is extensively evaluated using both synthetic signals and our ENF-WHU dataset consisting of $130$ real-world audio recordings, demonstrating substantially improved capability of extracting the ENF from realistically noisy observations over the existing single- and multi-tone competitors. This work further improves the applicability of the ENF as a forensic criterion in real-world situations.
\end{abstract}


\IEEEpeerreviewmaketitle

\section{Introduction}
\IEEEPARstart{T}{he} past decade has witnessed the active research and development progress of electric network frequency (ENF)-based multimedia forensics, enabled by the ENF properties of random fluctuation around the nominal frequency ($50$ or $60$ Hz) and intra-grid fluctuation consistency \cite{Experiment_General}. It was originally discovered that the ENF could be captured by AC-powered audio recording devices or DC-powered ones in the proximity of power mains \cite{Experiment_General}, while it has been subsequently discovered that the ENF could also exist in video recordings \cite{Video_Seeing_ENF} as well as in a single image \cite{App_Geo_Image}. Therefore, the time-dependent unique ENF pattern could be extracted from a testing multimedia file and used as a special watermark for forensic analysis, with or without a reference ENF. The reference ENF data could be collected from a power socket using a transformer and a soundcard, or from an infrastructure like the wide-area power system frequency monitoring network (FNET) \cite{TPS_FNET,App_FNET_2016,Experiment_PowerDel}, or from the powerline electromagnetic radiation (EMR) \cite{App_EMR}. 

Based on the ENF criterion, one could conduct forensic examinations including time-of-recording verification \cite{Experiment_General,Experiment_PowerDel,Own_DMA,App_BSim,Own_Matching_Error,App_ENF_Video_Stable_2020}, content forgery detection and localization \cite{Tamper_Phase,Tamper_Phase2,Tamper_IF,Tamper_Esprit_Hilbert,Own_AEM,Tamper_Offsets}, camera forensics \cite{App_Camera_Forensics}, region of recording estimation (geolocation) \cite{App_Geo_Image,App_Region1,App_Region2}, and others \cite{App_Counter_Measure,App_Database,App_Recapture,App_Gaussian_Model,Estimation_Compare}. However, prior to these applications, one has to extract a reliable ENF pattern from the testing file, which leads to the research topic of ENF estimation in audio recordings \cite{Estimation_IAA,Estimation_ImproveDFT,Estimation_Demodulation,Estimation_Harmonics,Estimation_Harmonics2,Estimation_LP}, while a few works are dedicated to a more challenging task of ENF extraction from videos \cite{Video_Seeing_ENF,Video_ENFVidDet,Video_Rolling_Shutter,App_ENF_Video_Stable_2020}. Some of the existing works on ENF-based forensic analysis have also investigated a few practical issues \cite{Own_Practical_Issue}, such as the analysis of ENF matching errors during time-of-recording verification \cite{Own_Matching_Error},  the study of factors affecting ENF capture \cite{App_Factor_Audio}, and reliable ENF extraction from non-static video recordings \cite{App_ENF_Video_Stable_2020}.  

Despite the existing research efforts on ENF-based multimedia forensics, there still exist several challenges when putting a solution into practical scenarios, among which the most significant ones would be \textbf{i)} the loss of ENF uniqueness when the duration of recording is relatively short and \textbf{ii)} the nature of the ENF being much weaker compared to noise and audio content in a recording. The first challenge is more precisely described as a phenomenon of the asymptotic property of a random signal when the observation is insufficient. It could be seen from most of the existing works that the durations of recordings in their experiments are at least longer than $10$ mins \cite{Video_Seeing_ENF,Experiment_PowerDel,App_BSim,Estimation_IAA,Estimation_Demodulation,Estimation_Harmonics,Estimation_Harmonics2,Estimation_LP}. For the second challenge, since the ENF nominal frequency (and several harmonics) is in very low frequency band, normal movement of the recording device could cause strong Doppler effects and destroy the ENF signal in audio \cite{App_Factor_Audio}. Besides, the ENF strength is inversely proportional to the squared distance between the microphone and the nearest ENF source, making it naturally weak when being captured. Unfortunately, more attentions have been paid to improved frequency estimators rather than noise and interference control, favoring moderate to high signal-to-noise ratio (SNR) situations, opposite to realistic ones.

In this paper, we focus on the problem of ENF estimation from real-world audio recordings and propose a robust framework as the solution. The proposed framework exploits two powerful means, i.e., ENF signal enhancement technique and multi-tone harmonic processing, to fulfill the purpose of extracting the ENF from very noisy observations. First, we extend the recently proposed robust filtering algorithm (RFA)\cite{Own_Singletone_Enhancement}, a single-tone ENF signal enhancement method, to the multi-tone scenario, and propose a harmonic robust filtering algorithm (HRFA), showing that the ENF harmonic components could respectively be enhanced without cross-component interference. The testing audio signal after ENF harmonic enhancement is then used for multi-tone model based ENF estimation. However, due to the weakness of the ENF signal and uncontrollable recording environments, some harmonic components could still be severely corrupted even after enhancement. In this situation, the use of corrupted harmonic components would disturb rather than facilitate ENF estimation. To solve this problem, we further propose an algorithm to find the optimal combination of harmonic components for ENF estimation, which is termed as graph-based harmonic selection algorithm (GHSA). In the GHSA, the harmonic selection problem is equivalently formulated as a maximum weight clique (MWC) problem, a classical problem in graph theory which is then solved with the use of the Bron-Kerbosch algorithm (BKA) \cite{MWC_Tutorial}. In this way, we obtain a subset of ENF harmonic components having maximized overall mutual correlation coefficients (CCs). Finally, the state-of-the-art multi-tone ENF estimators, i.e., maximum likelihood estimator (MLE) \cite{Estimation_Harmonics} and weighted MLE (WMLE) \cite{Estimation_Harmonics2} are incorporated in this paper for the final ENF estimators. The contributions of the paper are summarized as follows.
\begin{itemize}
	\item We propose the HRFA for ENF harmonic enhancement without cross-component interference, thus effectively improving the SNR conditions at each harmonic.
	\item We further propose the GHSA for the optimal selection of a subset of the ENF harmonic components, effectively ruling out severely corrupted components that deteriorate ENF estimation accuracy.
	\item With the above contributions, we construct to date the most robust ENF estimation framework consolidating three powerful tools, i.e., ENF harmonic enhancement, multi-tone harmonic processing, and the state-of-the-art MLE-based estimators.
\end{itemize}

The paper is organized as follows. Section II describes the proposed general framework and introduces the standalone MLE and WMLE. In Section III, we present the details of the proposed HRFA and GHSA. The performance of the proposed framework is extensively evaluated in Section IV using both synthetic data and real-world recordings, in which the state-of-the-art multi-tone ENF estimators are implemented for comparison. Conclusion is made in Section V.

\section{The Proposed General Framework}
In this section, we first present the multi-tone harmonic signal model and the proposed general framework for robust ENF estimation. Then we introduce and discuss about the incorporated state-of-the-art multi-tone ENF estimator, i.e., MLE \cite{Estimation_Harmonics} and WMLE \cite{Estimation_Harmonics2}.

\subsection{Signal Model}
We model the downsampled testing audio signal as the superposition of the ENF signal $\tilde{s}[n]$, audio content $\tilde{c}[n]$, and background noise $\tilde{v}[n]\sim\mathcal{N}(0,\sigma_{\tilde{v}}^2)$, given by
\begin{equation}
\tilde x[n] = \tilde s[n] + \tilde c[n] + \tilde v[n],
\end{equation}
and the SNR is defined as a function of $\tilde{s}[n]$ and $\tilde{v}[n]$,
\begin{equation}\label{SNR}
\operatorname{SNR} = \frac{\sum_n \tilde s^2[n]}{\sum_n \tilde v^2[n]}.
\end{equation}
The ENF signal $\tilde{s}[n]$ follows the multi-tone harmonic model consisting of $M$ harmonic components,
\begin{align}
\tilde s[n] & =\sum\limits_{m \in {\cal M}} {{A_m}[n]\cos \left( {2\pi T\sum\limits_{i = 0}^n {{f_m}[i]}  + {\phi _m}} \right)}  \notag\\
& =\sum\limits_{m \in {\cal M}} {{A_m}[n]\cos \left( {2\pi T\sum\limits_{i = 0}^n {mf[i]}  + {\phi}_m} \right)},\label{original_IF_model}
\end{align}
where $n\in\{0,1,\ldots,N-1\}$, $A_m[n]>0$ and $\phi_m$ are the unknown time-varying amplitudes and initial phase of the $m$th harmonic component respectively, $\mathcal{M}\subseteq \mathbb{Z}_+$ is the set of harmonic indices, $f[n]$ is the fundamental instantaneous frequency (IF), i.e., the ENF, measured in Hz, and $f_m[n]=mf[n]$ represents that the harmonic components are integer multiples of the fundamental component. Here $T=1/f_\text{S}$ is the sampling interval, and $f_\text{S}$ is the sampling frequency.

It should be noted that many existing consumer audio recording hardware devices and software packages cut off frequencies below $100$ Hz\footnote{For example, Sony ICD-TX650 digital recorder has a frequency response range of $[95, 20000]$ Hz, while Tascam DR-07X optionally cuts off frequencies below $40$, $80$, or $120$ Hz. Many smartphone recording apps cut off low frequencies even without letting the user know or providing an option.}, and as a consequence it is common that the fundamental ENF component is unavailable in a testing recording. Therefore, one usually resorts to a single higher harmonic component for ENF estimation, e.g., \cite{Own_AEM,Estimation_IAA,Estimation_Demodulation,Estimation_LP}. Considering such a real-world situation, we omit the fundamental component and normalize the estimated ENF to its $2$nd harmonic frequency band where necessary.

\subsection{General Framework and Workflow}
\begin{figure}[!t]
	\centering
	\includegraphics[width = 3.2in]{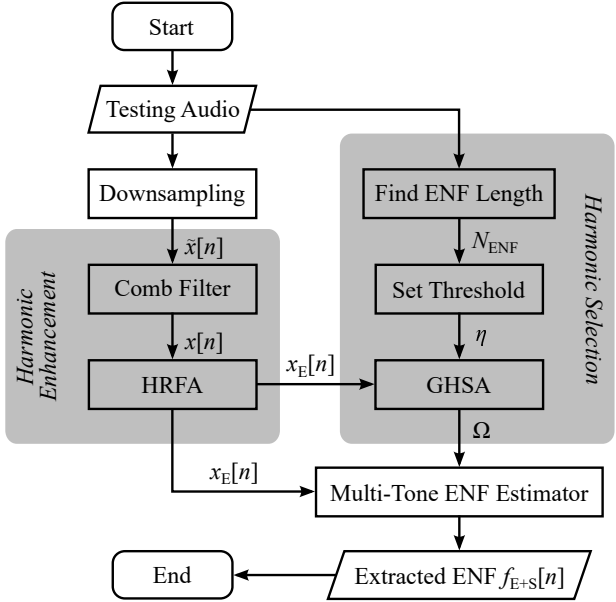}
	\caption{Flowchart of the proposed general framework.}
	\label{flowchart}
\end{figure}

\begin{figure}[!t]
	\centering
	\includegraphics[width=3.3in]{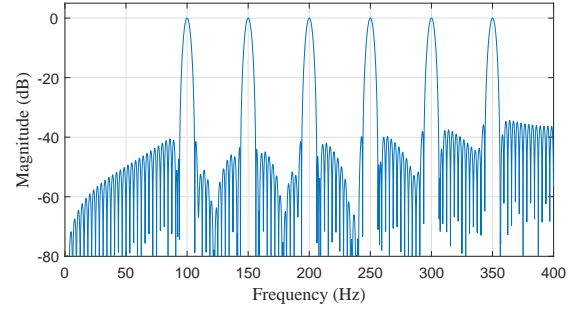}
	\caption{An example of the comb filter magnitude response for ENF harmonic bandpass filtering, where filter length is $256$, $|\mathcal{M}|=6$, and $f_\text{S}=800$ Hz.}
	\label{comb_filter}
\end{figure}

The flowchart of the proposed framework for robust ENF estimation is shown in Fig. \ref{flowchart}, where the gray parts exhibit the proposed modules for ENF harmonic enhancement and harmonic selection respectively. The workflow of the framework is as follows. The testing audio recording is first downsampled to a lower sampling frequency $f_\text{S}$ for computational efficiency. After that, a comb filter with $|\mathcal{M}|$ passbands is applied to the downsampled signal, filtering out the out-of-band noise at each targeted harmonic frequency band. Fig. \ref{comb_filter} depicts the magnitude response of a comb filter for $\mathcal{M}=\{2,3,4,5,6,7\}$. For the ease of notation, we remove the symbol $\{\tilde{\cdot}\}$ to denote the signals after bandpass filtering. We further assume that the ENF signal has been preserved, i.e., $s[n]=\tilde{s}[n]$, while the audio content has been filtered out. Therefore, the output of the comb filter is given by
\begin{equation}\label{filtered}
x[n] = s[n] + v[n] = \tilde{s}[n]+v[n],
\end{equation} 
which is then fed into the proposed HRFA for further harmonic enhancement. At the same time, according to the length of the recording, we could obtain the length of the ENF, denoted by $N_\text{ENF}$, and use it to generate an empirical threshold value $\eta$. The enhanced signal $x_\text{E}[n]$ and $\eta$ are fed into the GHSA which generates the optimal selection of ENF harmonic components, defined as a set of harmonic indices $\Omega\subseteq \mathcal{M}$. Finally $x_\text{E}[n]$ and $\Omega$ are passed to a multi-tone ENF estimator to yield the final estimation result.

The details of the proposed harmonic enhancement and harmonic selection algorithms will be provided in Section III. Before we proceed, we first review and comment on the state-of-the-art multi-tone model based ENF estimators, i.e., the MLE by Bykhovsky and Cohen \cite{Estimation_Harmonics} and the WMLE by Hajj-Ahmad \emph{et al.} \cite{Estimation_Harmonics2}, which are incorporated as a building block of the proposed framework.

\subsection{Performance Analysis of the MLE and WMLE}
Directly estimating $f[n]$ based on the model in (\ref{original_IF_model}) is very difficult due to its time-varying nature. Therefore, an approximated overlapping-frame-based model has been commonly considered, assuming a piecewise-constant IF per frame. Denote the frame length by $N_\text{F}$, then for the $l$th frame we have
\begin{align}
x_l[n] & = s_l[n]+v_l[n]\notag\\
& = \sum\limits_{m \in {\cal M}} {{A_{m,l}}[n]\cos \left( {2\pi Tmf[l]n + {\phi _{m,l}}} \right)} + v_l[n],\label{frame_model}
\end{align}
where $n\in\{0,1,\ldots,N_\text{F}-1\}$, $l\in\{0,1,\ldots,N_\text{ENF}-1\}$ and $f[l]$ is the constant ENF value within the frame. Note that both the MLE and WMLE are based on frame-based processing and the piecewise-constant assumption, and the workflow of the two estimators is in fact the flowchart in Fig. \ref{flowchart} omitting the gray components. 

The MLE of the fundamental frequency of a harmonic signal has been shown to be asymptotically optimal, i.e., it achieves the Cramer-Rao lower bound (CRLB) when $N_\text{F}$ approaches infinity \cite{Harmonic_CRLB,Estimation_Harmonics}. The MLE is implemented in \cite{Estimation_Harmonics} via the search within the periodogram for the maximum sum-of-squares, i.e., 
\begin{equation}
{f_{{\text{MLE}}}}[l] = \arg \mathop {\max }\limits_f \sum\limits_{m \in {\cal M}} {P_l(mf)},\label{MLE}
\end{equation}
where
\begin{equation}
P_l(f) = {\left| {\sum\limits_{n = 0}^{{N_{\text{F}}} - 1} {{x_l}[n]} \exp \left\{ { - \jmath 2\pi Tfn} \right\}} \right|^2}
\end{equation}
is the periodogram of $x_l[n]$. It can be seen from (\ref{MLE}) that all the harmonic components contribute to the estimation of the IF. The MLE is shown to achieve substantial performance improvements against the single-tone counterpart estimator especially in moderate to high SNR region.

A further step has been made in \cite{Estimation_Harmonics2} to cope with non-ideal SNR conditions, which introduces a set of weights defined as the estimated local SNR values. To estimate the local SNR values, each harmonic band is partitioned into a signal subband and a noise subband respectively \cite{Estimation_Harmonics2}, e.g., for the fundamental frequency band, the signal subband is $[49.98, 50.02]$ Hz and the noise subband is $[49, 49.98]\cup[50.02,51]$ Hz, while for the harmonics, the corresponding subbands are linearly scaled by the harmonic indices. The weights are simply the energy in signal subband divided by the energy in noise subband, and a larger weight indicates higher local SNR and more contribution in estimating the ENF. Note that the same approach has also been incorporated in \cite{Tamper_Esprit_Hilbert}. Denote the weight for the $m$th harmonic in the $l$th frame by $w_{m,l}$, then the WMLE searches for the maximum of sum-of-weighted-squares, i.e.,
\begin{equation}
{f_{{\text{WMLE}}}}[l] = \arg \mathop {\max }\limits_f \sum\limits_{m \in {\cal M}} w_{m,l}{P_l(mf)}.\label{WMLE}
\end{equation}

\begin{figure}[!t]
	\centering
	\includegraphics[width=3.35in]{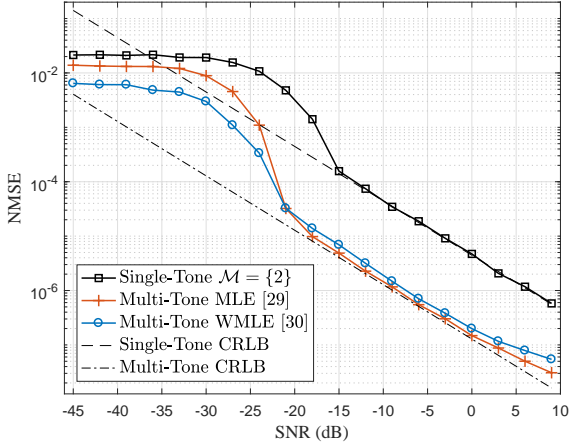}
	\caption{Theoretical analysis of the MLE and WMLE for IF estimation, where $\mathcal{M}=\{2,3,4,5,6,7\}$ for multi-tone estimators and $\mathcal{M}=\{2\}$ for the single-tone counterpart. Results are obtained via $10^3$ Monte Carlo experiments, where $N_\text{F}=8f_\text{S}$, NMSE calculated at the $2$nd harmonic.}
	\label{MSE_CRLB}
\end{figure}

We now compare and comment on the performances of the two estimators. The CRLB for an unbiased multi-tone harmonic frequency estimator is given by \cite{Harmonic_CRLB}
\begin{equation}
\operatorname{CRLB} \approx \frac{{24\sigma _{\tilde v}^2}}{{N_{\text{F}}^3}}{\left( {\sum\limits_{m \in {\cal M}} {\sum\limits_{n = 0}^{{N_{\text{F}}} - 1} {\frac{{{m^2}A_m^2[n]}}{{{N_{\text{F}}}}}} } } \right)^{ - 1}}{\left( {\frac{{{f_{\text{S}}}}}{{2\pi }}} \right)^2}, \label{CRLB}
\end{equation}
measured in Hz$^2$, while the CRLB for a single-tone estimator could be obtained by setting $|\mathcal{M}|=1$ in (\ref{CRLB}). The theoretical performances of the MLE and WMLE measured by normalized mean squared error (NMSE) are presented in Fig. \ref{MSE_CRLB}, based on $1000$ Monte Carlo realizations of white Gaussian noise (WGN), where the single-tone result is obtained by setting $\mathcal{M}=\{2\}$ in (\ref{MLE}). The CRLB curves are obtained by assuming a time- and harmonic- invariant amplitude value, i.e., $\forall m,n$, $A_m[n]=A$, thus it follows from (\ref{CRLB}) that
\begin{align}
\operatorname{CRLB} {{{|}}_{A,2{\text{nd}}}} & = \frac{{24\sigma _{\tilde v}^2}}{{N_{\text{F}}^3{A^2}}}{\left( {\sum\limits_{m \in \cal{M}} {{m^2}} } \right)^{ - 1}}{\left( {\frac{{{f_{\text{S}}}}}{{2\pi }}} \right)^2} \times 4\notag\\
& = \frac{{72}}{{N_{\text{F}}^3\operatorname{SNR} }}{\left( {\sum\limits_{m \in\cal{M}} {{m^2}} } \right)^{ - 1}}{\left( {\frac{{{f_{\text{S}}}}}{{2\pi }}} \right)^2} \times 4,
\end{align}
where the SNR defined in (\ref{SNR}) is equivalently expressed by the power ratio $6({A^2}/2\sigma _{\tilde v}^2)$, having $6$ harmonic components, and the scalar $4$ normalizes the NMSE values from the fundamental to the $2$nd harmonic. 
In this figure, a clear theoretical performance gain is observed by switching from the single-tone model to the multi-tone one. It can also be seen that the NMSE curves become flat in very low SNR region. This is because the search regions are bounded by the predefined frequency bands, and the estimated IFs could not be outside of the range $m\times [49.9, 50.1]$ Hz. In low SNR region the WMLE is seen to withstand the noise better while in moderate to high SNR region the NMSEs of the MLE are closer to the CRLB. 

The limitations of using the MLE or WMLE alone for ENF estimation and the corresponding motivations of our proposals are summarized as follows.
\begin{itemize}
	\item The moderate to high SNR condition for the estimators to achieve satisfactory performances is very likely to be invalid in real-world audio recordings which contain strong audio content and non-stationary noise. This calls for more attentions on noise control rather than the estimator, which motivates us to develop the HRFA.
	\item In real-world audio recordings, some harmonic components may be severely corrupted and unusable even after enhancement. The corrupted harmonics would disturb rather than facilitate IF estimation. In such a common situation, it is more important to appropriately select a subset of harmonic components than increasing the number of the harmonics without knowing the quality. This motivates us to develop the GHSA.
\end{itemize}

\section{The Proposed Algorithms}
\subsection{The Harmonic Robust Filtering Algorithm}
Recently, we have proposed the RFA \cite{Own_Singletone_Enhancement} for single-tome ENF signal enhancement, which encodes $x[n]$ as the IFs of a complex analytical sinusoidal frequency modulated (SFM) signal $z[n]$, followed by the use of a kernel function whose averaged phase term yields the enhanced signal $x_\text{E}[n]$. For conciseness and to emphasize the contribution of this paper, we present only the essential components of the RFA here and focus more on how the single-tone method is extended to the multi-tone scenario, leading to the design of the HRFA. As shown in Fig. \ref{flowchart}, the input of the HRFA is $x[n]$ in (\ref{filtered}), i.e., the audio signal after downsampling and bandpass filtering. Note that the RFA and HRFA share the same input, but they process the input differently (The RFA only focuses on and processes the $2$nd harmonic). 

\subsubsection{The RFA \cite{Own_Singletone_Enhancement}}
The complex analytical SFM signal $z[n]$ is given by 
\begin{equation}\label{SFM}
z[n] = \exp\left\{ \jmath 2 \pi T \alpha \sum\limits_{i=0}^{n} x[i] \right\},
\end{equation}
where $x[n]$ becomes the IF of $z[n]$, and $\alpha>0$ controls the position of $x[n]$ in the frequency spectrum of $z[n]$. Define the instantaneous auto-correlation function of $z[n]$ as
\begin{equation}
{R}_z[n,\tau ] \triangleq z\left[ {n + \tau } \right]{z^*}\left[ {n - \tau } \right],
\end{equation} 
where $\{\cdot\}^*$ denotes complex conjugation and $\tau$ is the lag parameter, then the kernel function is given by
\begin{equation}
K\left[ {n,\tau } \right] = R_z{[n,\tau ]^{\theta_0 \sin (2\theta_0) }}R_z{[n,\tau']^{\theta_0 \cos(2\theta_0) }},\label{kernel}
\end{equation}
where
\begin{equation}
\theta_0 \buildrel \Delta \over =  \pi Tf[n]\tau,
\end{equation}
$\tau'=\tau+\left[\kern-0.15em\left[ {{f_{\text{S}}}/4f[n]} \right]\kern-0.15em\right]$, $\left[\kern-0.15em\left[\cdot \right]\kern-0.15em\right]$ is the rounding operator. Here $f[n]$ is the probing IF, and it provides an initial guess and is updated iteratively in the RFA. It has then been proven that the averaged phase of the kernel function yields the denoised output, i.e.,
\begin{equation}\label{averaged_phase}
x_\text{E}[n] \approx s[n] = \frac{{\sum\nolimits_{i =  0 }^{ \tau } {\Theta \{ {K}[n,i]\} } }}{{\left( {\tau  + 1} \right)\tau\pi \alpha }},
\end{equation}
where $\Theta\{\cdot\}$ is the phase operator. The above process uses the nominal value of the $2$nd harmonic as the initial guess, i.e., $\forall n$, $f[n]=100$, which is substituted into (\ref{kernel}). Once $x_\text{E}[n]$ is obtained via (\ref{averaged_phase}), a single-tone ENF estimator is applied to update the value of $f[n]$ which is then substituted into (\ref{kernel}) again. Such a process is repeated for $I$ iterations.

\subsubsection{The Proposed HRFA}
The RFA was originally proposed to enhance an arbitrary multi-component signal contaminated by additive noise \cite{Enhancement_RTVF}, which was then shown to be specially suitable and powerful to enhance the ENF signal in the single-tone scenario \cite{Own_Singletone_Enhancement}. This is because the RFA achieves the optimal enhancement performance if the error between the initial guess of the IFs and the ground truth could be tightly bounded, while the ENF satisfies this condition exactly, i.e., its dynamic range is less than $1$ Hz at the $2$nd harmonic and the initial guess of a constant $100$ Hz would be a sound setting for the algorithm to perform its task. Theoretical and quantitative analysis of this condition is provided in \cite{Own_Singletone_Enhancement} and is not presented here. 

To develop the HRFA, we further exploit another important property of the RFA, which is stated as if the components of the signal are separated in time-frequency domain, then it is possible for the RFA to respectively enhance each component without cross-component interference. Detailed analysis of this condition could be found in \cite{Enhancement_RTVF}, while here we show that this condition holds for the harmonic model (\ref{original_IF_model}), a special case of multi-component models, under a reasonable number of harmonics. Using the loose subbands mentioned in Section II-C, the $m$th ENF harmonic component is bounded within $m\times [49, 51]$ Hz. Let the frequency distance between the $(m-1)$th and $m$th components be $d(m)$ Hz, then it follows that
\begin{equation}
d(m) \ge 49m - 51(m - 1) = 51 - 2m,
\end{equation}
indicating that it is only possible for the $26$th or higher ($m \ge 26$) harmonic components to overlap with their neighbors, while lower harmonic components are strictly and sufficiently separated. This verifies the feasibility of applying the RFA to the harmonic scenario. 

The problem now reduces to finding a sound initial guess of the IFs for each harmonic component. Intuitively one may use the nominal values $m \times 50, \forall m\in \mathcal{M}$. However, due to the scaling effect among the harmonics, the errors between the initial guess and the ground truth are up-scaled by $m$. As a result, higher order harmonics are prone to more performance degradation. To avoid this problem, the HRFA is designed in such a way that the initial guesses of other harmonic components are the linearly scaled versions of the enhanced $2$nd one. The reasons of doing so are \textbf{i)} the fundamental component is usually unavailable in real-world recordings, thus the $2$nd harmonic component has the minimum initial guess error among all available harmonic candidates; \textbf{ii)} the use of the scaled versions of the enhanced $2$nd harmonic component instead of the constant nominal values as the initial guesses can effectively minimize the initial errors in the respective harmonic components. 

Based on the above analysis, the HRFA is described as follows. Denote the time-domain expression of the $m$th harmonic component of $x[n]$ by $\{x[n]\}_m$ whose frequencies are bound by $m\times [49.9, 50.1]$ Hz, then we can rewrite (\ref{filtered}) as
\begin{equation}
x[n] = \sum\limits_{m \in {\cal M}} {{{\{ x[n]\} }_m}},
\end{equation} 
consisting of the signal components and in-band noise respectively. The enhanced components are then denoted by $\{ x_\text{E}[n]\}_m$. Since we now consider multiple components rather than a single one in the original RFA, the kernel function in (\ref{kernel}) becomes component-dependent, which is modified to
\begin{align}
& K\left[ n, \tau, m \right] =\notag\\
& \quad R_z{[n,\tau ]^{\theta(m) \sin [2\theta(m)] }}R_z{[n,\tau']^{\theta(m) \cos[2\theta(m)] }},\label{kernel_multi}
\end{align}
where
\begin{equation}
\theta(m) \buildrel \Delta \over =  \pi T m f[n]\tau,
\end{equation}
and (\ref{averaged_phase}) becomes
\begin{equation}\label{averaged_phase_multi}
\{x_\text{E}[n]\}_m \approx \{s[n]\}_m = \frac{{\sum\nolimits_{i =  0 }^{ \tau } {\Theta \{ {K}[n,i,m]\} } }}{{\left( {\tau  + 1} \right)\tau\pi \alpha }}.
\end{equation}
Therefore, the output of the original RFA is in fact $\{x_\text{E}[n]\}_2$, a special case of the harmonic scenario. Note that (\ref{kernel_multi})\---(\ref{averaged_phase_multi}) are valid only if the initial guess errors are bounded and the harmonic components are sufficiently separated. Following the above-mentioned mechanism, the HRFA is summarized in \textbf{Algorithm 1}, where short-time Fourier transform (STFT) is used as the intermediate IF estimator. It can be seen that this algorithm first enhances and estimates the $2$nd harmonic component and then uses its scaled versions as the initial guesses of other components, while Steps $3$\---$12$ is equivalent to the original RFA \cite{Own_Singletone_Enhancement}. 

\IncMargin{.5em}
\begin{algorithm}[!t]
	\caption{The Proposed HRFA.}
	\setstretch{1}
	\KwIn{$x[n]$, $\alpha$, $\tau$, $I$, $\mathcal{M}$;}
	\KwOut{$x_\text{E}[n]$;}
	Initial guess of fundamental: $\forall n, f[n]\leftarrow 50$\;
	Encode $x[n]$ using (\ref{SFM}) and obtain $z[n]$\;
	\While{$i=1, i \le I$}
	{
		\For{$n=0$ \KwTo $N-1$}
		{			
			Update $K[n,\tau, 2]$ using (\ref{kernel_multi})\;
			Update $\{x_\text{E}[n]\}_2$ using (\ref{averaged_phase_multi})\;
		}
	    $f[l]\leftarrow$ STFT of $\{x_\text{E}[n]\}_2$ plus peak search\;
	    $f[n]$ $\leftarrow$ interpolate $f[l]$: $\mathbb{R}^{N_\text{FFT}\times 1}\rightarrow \mathbb{R}^{N\times 1}$\;
		$i++$\;
	}
    \For{$\forall m\in \mathcal{M}$, $m\neq 2$}
    {
    	\While{$i=1, i \le I$}
    	{
    		\For{$n=0$ \KwTo $N-1$}
    		{			
    			Update $K[n,\tau, m]$ using (\ref{kernel_multi})\;
    			Update $\{x_\text{E}[n]\}_m$ using (\ref{averaged_phase_multi})\;
    		}
    		$f[l]\leftarrow$ STFT of $\{x_\text{E}[n]\}_m$ plus peak search\;
    		$f[n]$ $\leftarrow$ interpolate $f[l]$: $\mathbb{R}^{N_\text{FFT}\times 1}\rightarrow
    		\mathbb{R}^{N\times 1}$\;
    		$i++$\;
    	}   
   }
   ${x_{\text{E}}}[n] \leftarrow  \sum\nolimits_{m \in {\cal M}} {{{\{ {x_{\text{E}}}[n]\} }_m}}$\;
\end{algorithm}
\DecMargin{.5em}

\subsection {The Graph-Based Harmonic Selection Algorithm}
\subsubsection{Harmonic Selection Problem}
Based on the output of the HRFA ${x_{\text{E}}}[n]$, one could respectively estimate the IFs in each harmonic frequency band, denoted by $\{\hat{f}[l]\}_m$. For the ease of notation, we drop the sign $\hat{\{\cdot\}}$ in the following derivations. 

\begin{figure}[!t]
	\centering
	\includegraphics[width=3.4in]{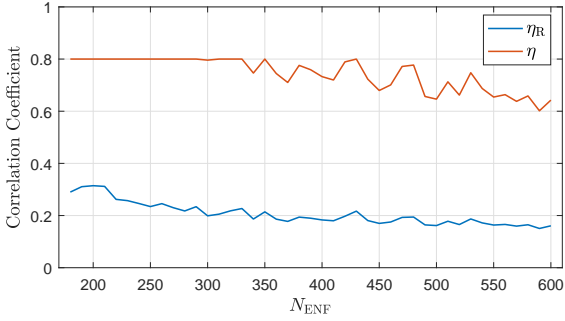}
	\caption{Empirical results of $\eta$, where $\kappa =4$ and $N_\text{R}=10^4$.}
	\label{threshold}
\end{figure}

Because of the uncontrollability of noise and interference and the natural weakness of the ENF, some harmonic components may still be severely corrupted after the HRFA. These corrupted components would disturb rather than facilitate the subsequent multi-tone estimator. Therefore, we need to select the most suitable subset of harmonic components to improve ENF estimation performance. We consider the set of selected harmonic indices $\Omega$ to be optimal if their averaged mutual CC is maximized. This optimization problem is formulated as follows,
\begin{subequations}\label{optimization}
\begin{align}
\mathop {{\text{maximize}}}\limits_{\Omega}& \quad{|\Omega| \choose 2}^{-1}\sum\limits_{m_i,m_j \in \Omega ,m_i \ne m_j} {\rho (m_i,m_j)} \label{cost}\\
{\text{subject to}}&\quad {\Omega  \subseteq {\cal M},}\\
& \quad {\rho \left(m_i,m_j\right) \ge \eta, \forall m_i,m_j \in \Omega ,m_i \ne m_j},\label{constraint}
\end{align}
\end{subequations}
where $n \choose k$ is the ``$n$ choose $k$'' binomial combination operator, $0\le\rho(\cdot,\cdot)\le 1$ is the CC operator, $\rho(m_i,m_j)$ is the simplified notation of $\rho \left(\{ f[l]\}_{m_i}, \{ f[l]\}_{m_j}\right)$, $m_i, m_j \in \mathcal{M}$, $i,j\in\{1,2,\ldots,|\mathcal{M}|\}$, and $\eta$ is an empirical threshold designed as a function of the length of the ENF $N_{\text{ENF}}$. Let the frame step-size be $\Delta$ samples and assume that $x[n]$ is appropriately zero-padded, then we have
\begin{equation}
N_\text{F} + (N_\text{ENF} - 1)\Delta = N \Rightarrow {N_{{\text{ENF}}}} = \frac{{N - {N_{\text{F}}}}}{\Delta } + 1.
\end{equation}
To obtain the threshold value $\eta$, we generate two independent pseudorandom WGN sequences of length $N_\text{ENF}$ and calculate their CC. This process is repeated for $N_\text{R}$ times and the maximum CC is selected as the reference value, denoted by $\eta_\text{R}$. Then we empirically set 
\begin{equation}\label{eta}
\eta = \operatorname{min}\{\kappa\eta_\text{R}, 0.8\},
\end{equation}
meaning that a CC is acceptable if it is greater than $0.8$ or at least $\kappa>1$ times greater than the maximum empirical CC of two independent WGN sequences of length $N_\text{ENF}$. Fig. \ref{threshold} shows the threshold value as a function of $N_\text{ENF}$, where $N_\text{R}=10^4$. According to the asymptotic property, increasing $N_\text{ENF}$ results in the decreasing of the CC between independent WGN sequences, thus $\eta$ has a decreasing trend. 

\subsubsection{Equivalent Formulation and Solution}The problem (\ref{optimization}) is generally a difficult integer programming problem. We now show that this problem is equivalent to the classical MWC problem in graph theory, which is then solved by the proposed GHSA. Note that the MWC problem has been formulated in diverse research fields including video analytics \cite{MWC_Video}, EEG signal processing \cite{MWC_EEG}, and communications \cite{MWC_Tutorial}, among others, while in this paper we make the first attempt to solve the multi-tone model based ENF estimation problem from the perspective of clique. 

\begin{figure}[!t]
	\centering
	\includegraphics[width=2.6in]{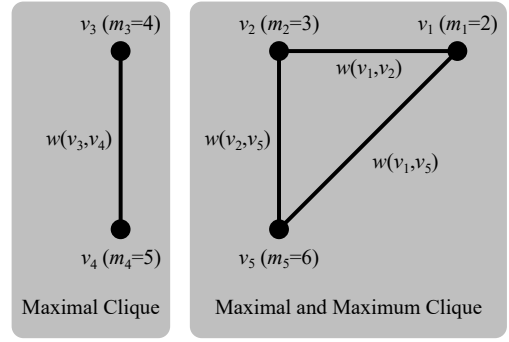}
	\caption{An example of the graph constructed by $\mathbf{R}_\text{adj}$, where $\mathcal{M}=\{2,3,4,5,6\}$. Vertex: $v_i$. Edge: $(v_i, v_j)$. Weight: $w(v_i,v_j)=\rho(m_i,m_j)$.}
	\label{graph_example}
\end{figure}

Let us construct the correlation matrix of the $|\mathcal{M}|$ harmonic IF sequences $\{{f}[l]\}_m$, denoted by $\mathbf{R}\in \mathbb{R}^{|\mathcal{M}|\times |\mathcal{M}|}$, whose elements are given by ${\{\mathbf{R}\}_{i,j}} = \rho \left(m_i,m_j\right)$, where $m_i, m_j \in \mathcal{M}$, and $i,j\in\{1,2,\ldots,|\mathcal{M}|\}$. Then, according to the constraint (\ref{constraint}), we set all elements which are less than $\eta$, as well as the diagonal ones to zero. Denote the $\mathbf{R}$ after this process by $\mathbf{R}_\text{adj}$. If $\mathbf{R}_\text{adj}$ is an all-zero matrix, then there does not exist a single pair of harmonic components with satisfactory correlation property. In this situation, based on the slowly varying nature of the ENF, the smoothest harmonic component is selected as the single component for ENF estimation, i.e.,
\begin{equation}\label{single_component}
\Omega  = \arg \mathop {\min }\limits_m \sum\limits_{l = 1}^{{N_{{\text{ENF}}}} - 1} {\left| {{{\{ f[l]\} }_m} - {{\{ f[l - 1]\} }_m}} \right|},
\end{equation}
where $m\in \mathcal{M}$, the smoothness is quantified by the sum-of-absolute-difference, and $|\Omega|=1$. If not all elements of $\mathbf{R}_\text{adj}$ are zero, then we could consider $\mathbf{R}_\text{adj}$ as the adjacency matrix of an edge-weighted undirected graph
$\mathcal{G}=(\mathcal{V},\mathcal{E})$, where $\mathcal{V}$ is the set of vertices equivalently mapped to $\mathcal{M}$, e.g., $\mathcal{M}=\{2,3,5\}\rightleftarrows\mathcal{V}=\{v_1,v_2,v_3\}$, and $\mathcal{E}\subseteq\{(v_i,v_j)|\forall v_i, v_j \in\mathcal{V}\}$ is the set of edges carrying weights using the nonzero elements of $\mathbf{R}_\text{adj}$, i.e., $w{(v_i,v_j)}=\rho(i,j)$. Before we proceed, the definitions of some important terms are provided as follows \cite{MWC_Tutorial}.
\begin{itemize}
	\item \emph{Clique:} a subset of vertices such that each vertex in the set is adjacent to all other vertices in the set, also known as a fully connected subgraph.
	\item \emph{Maximal clique:} a clique that cannot be extended by including more
	adjacent vertices, also known as a non-extensible fully connected subgraph.
	\item \emph{Maximum clique:} the maximal clique with the greatest
	number of vertices.
\end{itemize}
Fig. \ref{graph_example} shows an example of the graph constructed by $\mathbf{R}_\text{adj}$ whose elements indexed by $(1,3)$, $(1,4)$, $(2,3)$, $(2,4)$, $(3,5)$, and $(4,5)$, and the corresponding symmetric ones are all zero. The vertices are mapped to the harmonic indices. In this figure, there are two maximal cliques, i.e., $\{v_1, v_2, v_5\}$ and $\{v_3, v_4\}$, respectively, while the former is also the maximum clique.

The MWC further takes into consideration the weights of edges or vertices, which could be defined in different ways according to the specific problem under investigation \cite{MWC_Tutorial}. According to (\ref{cost}), we define the corresponding MWC as the maximal clique with the greatest averaged edge weight. Therefore, problem (\ref{optimization}) could be reformulated as the following MWC problem, 
\begin{subequations}\label{optimizationC}
\begin{align}
\mathop {{\text{maximize}}}\limits_{\mathcal{C}} & \quad {|\mathcal{C}| \choose 2}^{-1}\sum\limits_{{v_i},{v_j} \in \mathcal{C},i \ne j} {w({v_i},{v_j})} \label{costC}\\
{\text{subject to}} & \quad (v_i,v_j) \in \mathcal{E},\forall v_i,v_j \in \mathcal{C},\label{constraintC}
\end{align}
\end{subequations}
where (\ref{costC}) ensures maximum averaged weight, and constraint (\ref{constraintC}) states that all the edges of $\mathcal{C}$ belong to $\mathcal{E}$, i.e., $\mathcal{C}\subseteq \mathcal{V}$ is a clique. Note that there also exists a one-to-one mapping between $\mathcal{C}$ and $\Omega$, e.g., if $\mathcal{M}=\{2,3,4\}\rightleftarrows\mathcal{V}=\{v_1,v_2,v_3\}$ and $\mathcal{C}=\{v_1,v_3\}$, then $\Omega=\{2,4\}$.

\IncMargin{.5em}
\begin{algorithm}[!t]
	\caption{The Proposed GHSA.}
	\setstretch{1}
	\KwIn{$x_\text{E}[n]$, $\eta$, $\mathcal{M}$;}
	\KwOut{$\Omega$;}
	$\{f[l]\}_m$ $\leftarrow$ STFT of $x_\text{E}[n]$ plus peak search, $\forall m\in{\mathcal{M}}$\;
	$\rho(m_i,m_j)\leftarrow\rho \left(\{ f[l]\}_{m_i}, \{ f[l]\}_{m_j}\right)$, $\forall m_i,m_j\in\mathcal{M}$\;
	Obtain $\mathbf{R}$, ${\{\mathbf{R}\}_{i,j}} \leftarrow \rho \left(m_i,m_j\right)$, $\forall i,j\in\{1,\ldots,|\mathcal{M}|\}$\;
	$\mathbf{R}_{\rm{adj}} \leftarrow \mathbf{R}$\;
	$\{\mathbf{R}_{\rm{adj}}\}_{i,j}=0$, if $\{\mathbf{R}_{\rm{adj}}\}_{i,j}<\eta$ or $i=j$\; 
	\uIf{$\{\mathbf{R}_{\rm{adj}}\}_{i,j}=0$, $\forall i,j$}
	{Obain $\Omega$ using (\ref{single_component})\;}
	\Else{
		Construct graph $\mathcal{G}=(\mathcal{V},\mathcal{E})$ using ${\mathbf{R}_\text{adj}}$\;
		$\{\mathcal{C}_k\}$ $\leftarrow$ maximal cliques of $\mathcal{G}$ using the BKA\;
		Obtain $\mathcal{C}$ by substituting $\{\mathcal{C}_k\}$ into (\ref{MWC_solution})\;
		Map $\mathcal{C}$ to $\Omega$\;
	}
\end{algorithm}
\DecMargin{.5em}

Note that (\ref{optimizationC}) is slightly different from the existing MWC problem formulations due to the extra averaging term in (\ref{costC}). There thus may not be a directly applicable solution to the problem. Here, we propose a two-stage approach to solve (\ref{optimizationC}), which first finds all maximal cliques of $\mathcal{G}$ using the BKA \cite{MWC_Tutorial} and then selects the one having the greatest averaged edge weight. Specifically, assume $K$ maximal cliques are found by the BKA, which are denoted by a set $\{\mathcal{C}_k\}$, where $\forall k\in\{1,2,\ldots,K\}$, $\mathcal{C}_k \subseteq \mathcal{V}$, then the solution is given by
\begin{equation}\label{MWC_solution}
\mathcal{C} = \arg \mathop {\max }\limits_{\mathcal{C}_k}  {{|\mathcal{C}_k|}\choose 2}^{-1}\sum\limits_{{v_i},{v_j} \in {\mathcal{C}_k}, i\ne j} {w({v_i},{v_j})},
\end{equation}
which is then mapped to $\Omega$ containing the selected harmonic indices. We note here that a more efficient solution to (\ref{optimizationC}) may exist, but at the moment it is not in the scope of this paper. Based on the above process, the GHSA is summarized in \textbf{Algorithm 2}.

Recall Fig. \ref{flowchart}, using the enhanced signal $x_\text{E}[n]$ and the selected harmonic indices $\Omega$, the MLE or WMLE is applied to yield the final ENF estimation result. Take the MLE as an example, the proposed framework finally modifies (\ref{MLE}) into
\begin{equation}
{f_{{\text{P-MLE}}}}[l] = \arg \mathop {\max }\limits_f \sum\limits_{m \in {\Omega}} {P_{\text{E},l}(mf)},\label{MLE_omega}
\end{equation}
where 
\begin{equation}
P_{\text{E},l}(f) = {\left| {\sum\limits_{n = 0}^{{N_{\text{F}}} - 1} {{x_{\text{E},l}}[n]} \exp \left\{ { - \jmath 2\pi Tfn} \right\}} \right|^2}
\end{equation}
is the periodogram of the $l$th frame of the enhanced signal $x_\text{E}[n]$, denoted by $x_{\text{E},l}[n]$, and the harmonic index set is changed from $\mathcal{M}$ to $\Omega$.

\begin{table}[!t]
	\renewcommand{\arraystretch}{1.2}
	\setlength{\tabcolsep}{4pt}
	\caption[]{Summary of ENF estimation schemes for comparison}
	\label{estimator_summary}
	\centering
	\vspace*{-6pt}
	\begin{tabular}{c|l|c|c|c|c}
		\hline
		\hline
		Group & \multicolumn{1}{c|}{{Scheme}} & $|\mathcal{M}|$ & $x_\text{E}[n]$ & Selection & $\Omega$\\
		\hline
		\multirow{4}{*}{I}& $f_\text{single}[l]$ & $=1$ & No & No & $=2$ \\
		&$f_\text{E-single}[l]$ \cite{Own_Singletone_Enhancement} & $=1$ & RFA & No & $=2$\\
		&$f_\text{MLE}[l]$ \cite{Estimation_Harmonics}  & $>1$ & No & No & $=\mathcal{M}$ \\	
		&$f_\text{WMLE}[l]$ \cite{Estimation_Harmonics2} & $>1$ & No & No& $=\mathcal{M}$\\
		\hline
		\multirow{4}{*}{II}&$f_\text{E-MLE}[l]$ & $>1$ & HRFA & No & $=\mathcal{M}$\\
		&$f_\text{E-WMLE}[l]$ & $>1$ & HRFA & No & $=\mathcal{M}$ \\
		&$f_\text{S-MLE}[l]$ & $> 1$ & No & GHSA & $\subseteq\mathcal{M}$\\
		&$f_\text{S-WMLE}[l]$ & $> 1$ & No & GHSA & $\subseteq\mathcal{M}$ \\
		\hline
		\multirow{2}{*}{III}&$f_\text{P-MLE}[l]$ & $> 1$ & HRFA & GHSA & $\subseteq\mathcal{M}$\\
		&$f_\text{P-WMLE}[l]$ & $> 1$ & HRFA & GHSA & $\subseteq\mathcal{M}$\\
		\hline
		\hline
	\end{tabular}
\end{table}

\begin{figure*}[!t]
	\centering
	\subfigure[$2$nd.]{\includegraphics[width=1.15in]{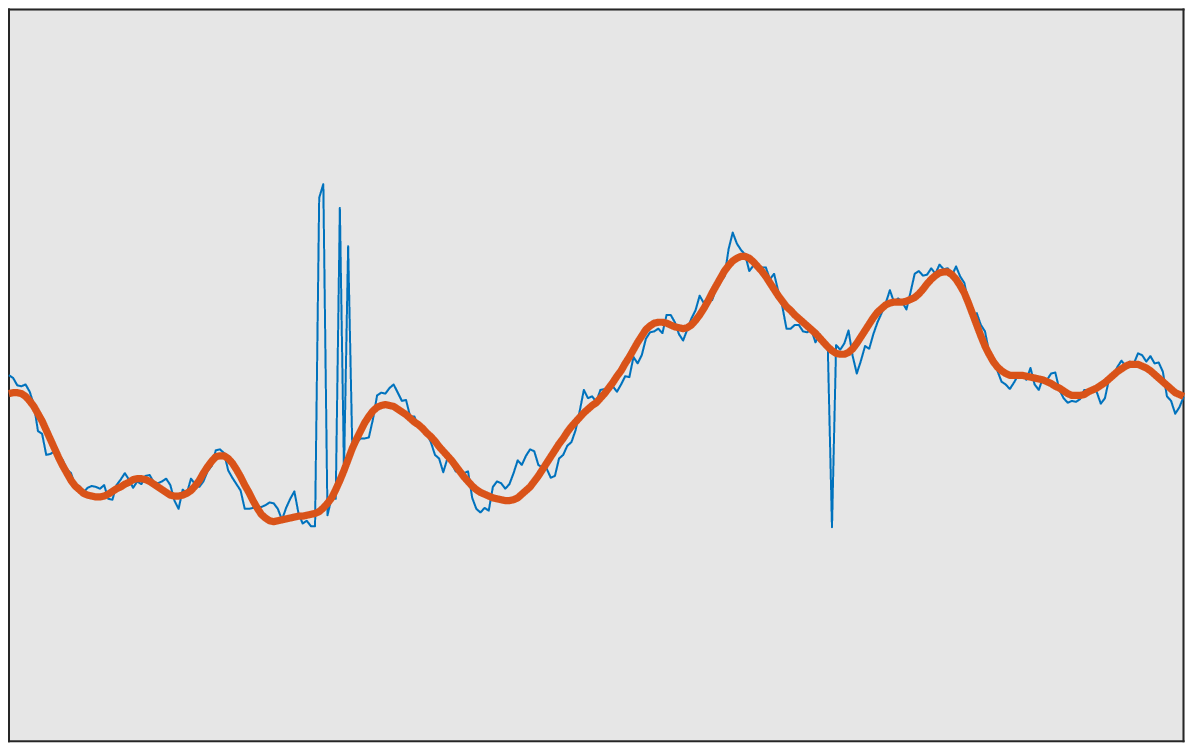}}
	\subfigure[$3$rd.]{\includegraphics[width=1.15in]{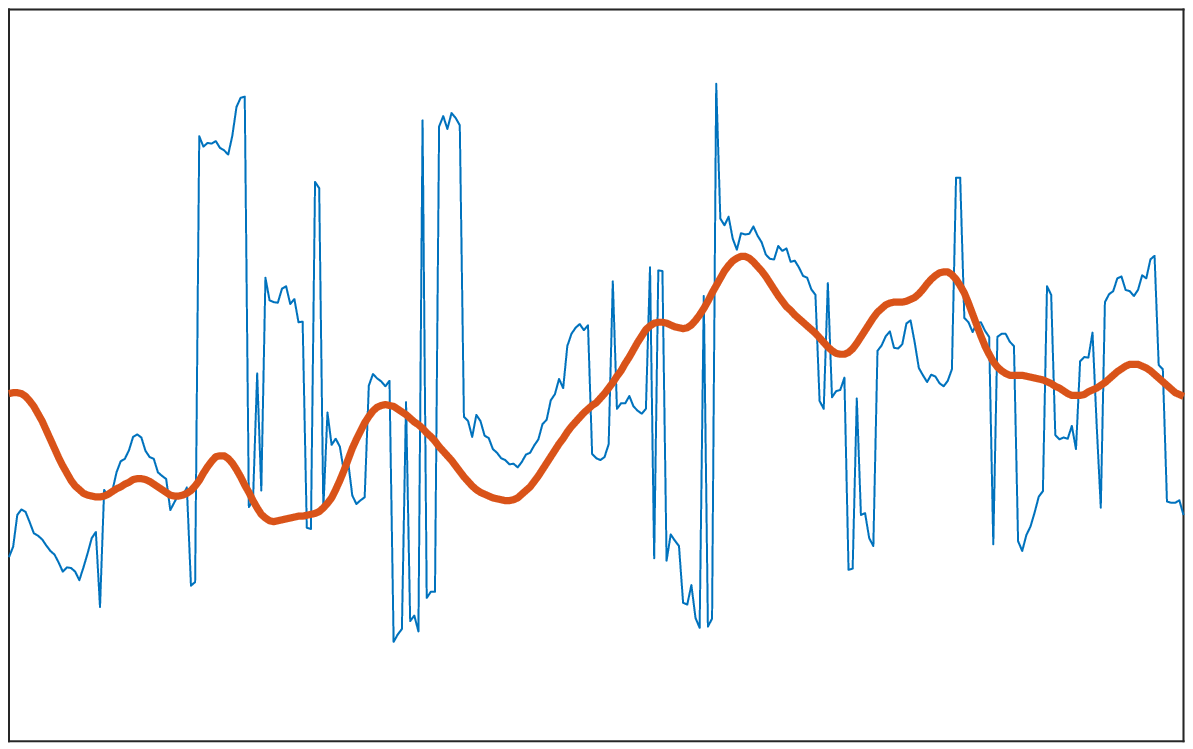}}
	\subfigure[$4$th.]{\includegraphics[width=1.15in]{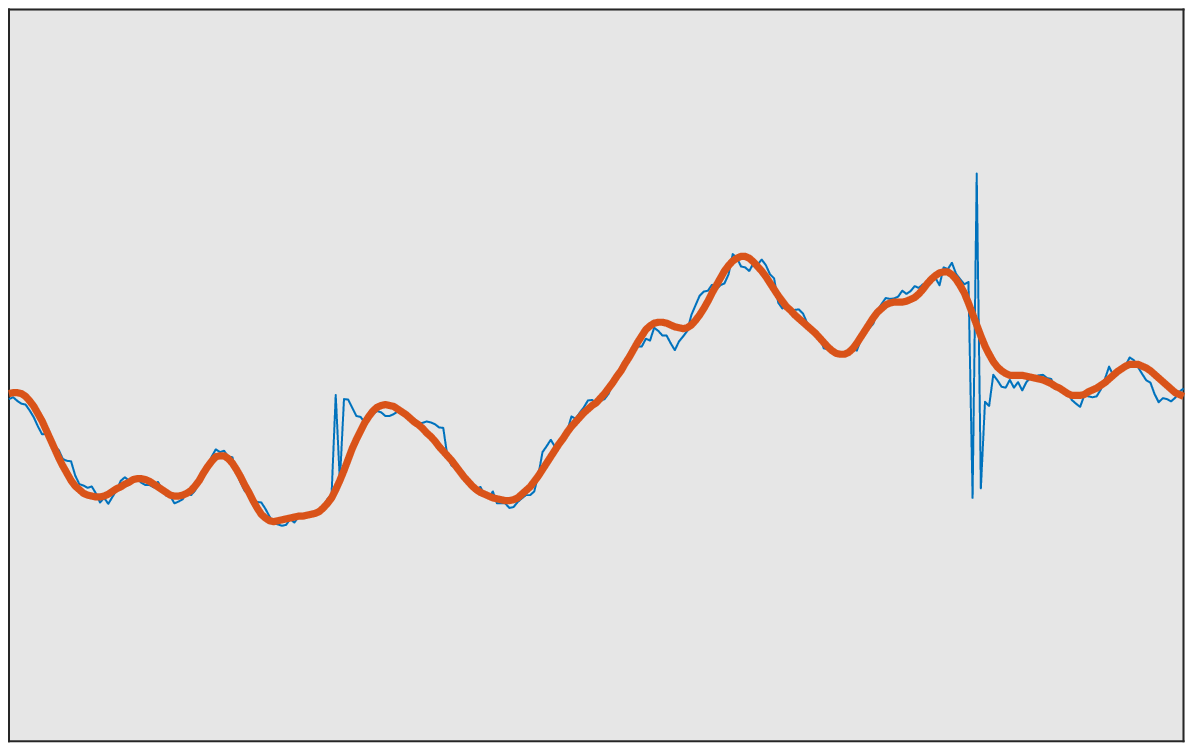}}
	\subfigure[$5$th.]{\includegraphics[width=1.15in]{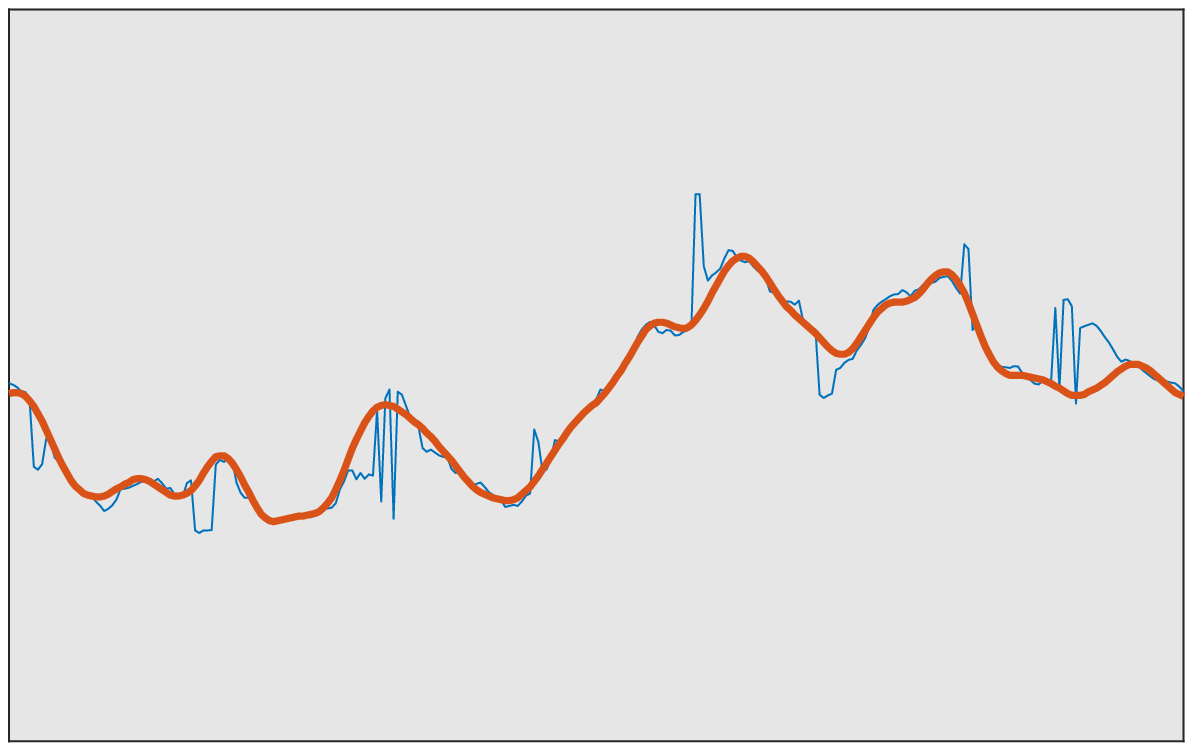}}
	\subfigure[$6$th.]{\includegraphics[width=1.15in]{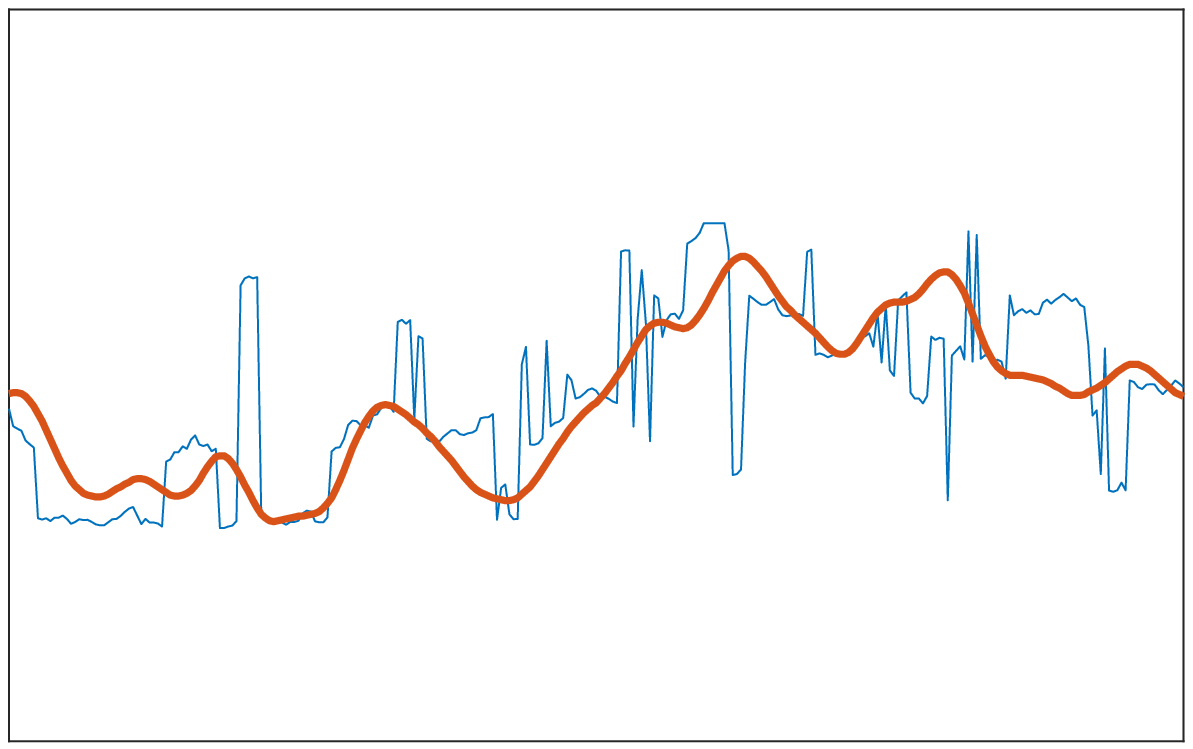}}
	\subfigure[$7$th.]{\includegraphics[width=1.15in]{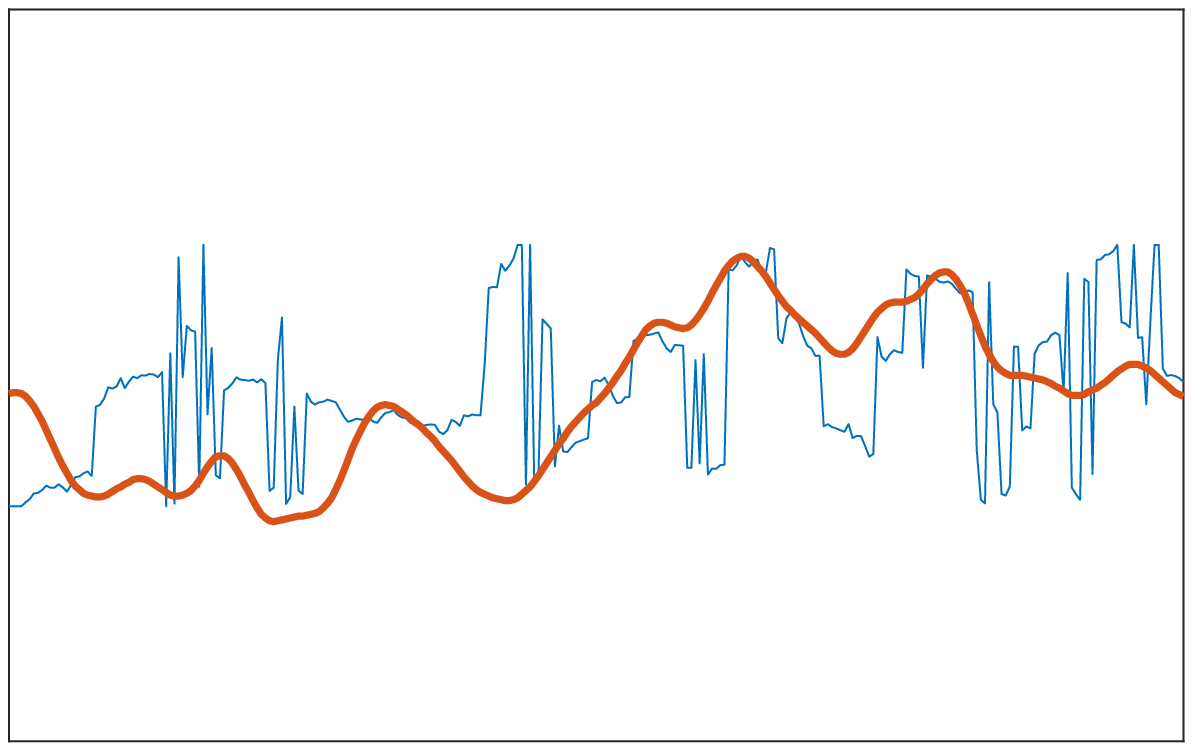}}\\
	\subfigure[$2$nd enhanced.]{\includegraphics[width=1.15in]{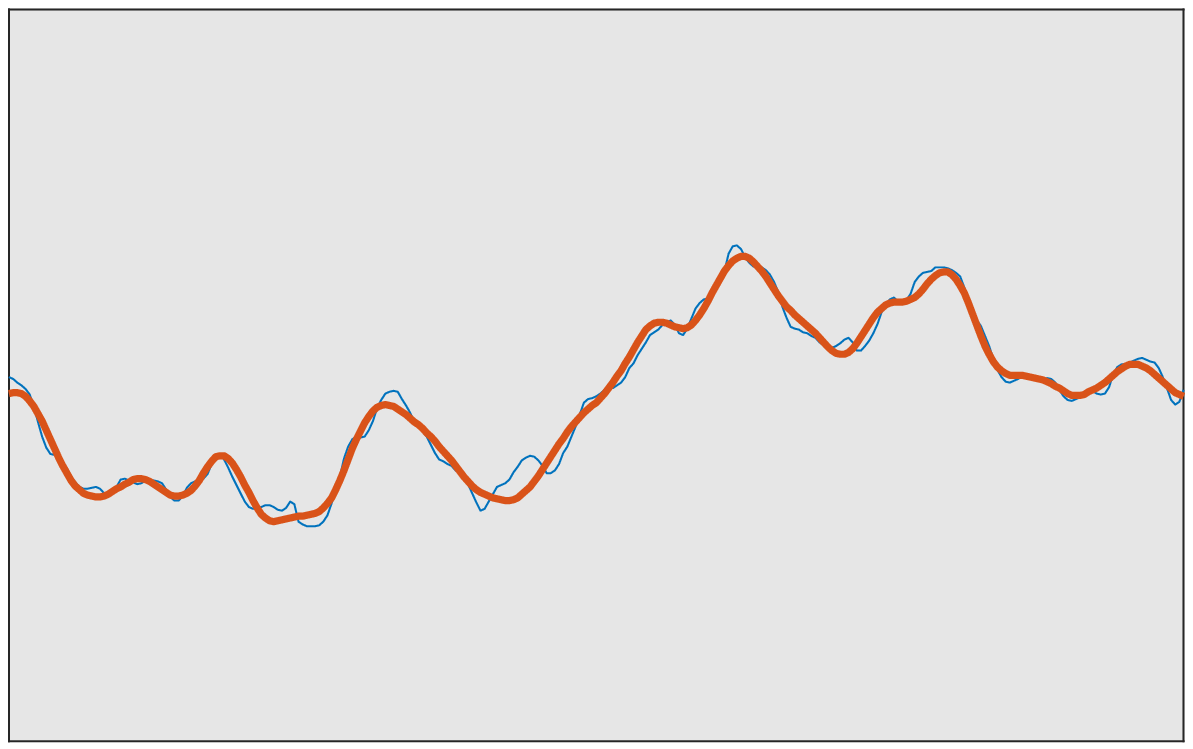}}
	\subfigure[$3$rd enhanced.]{\includegraphics[width=1.15in]{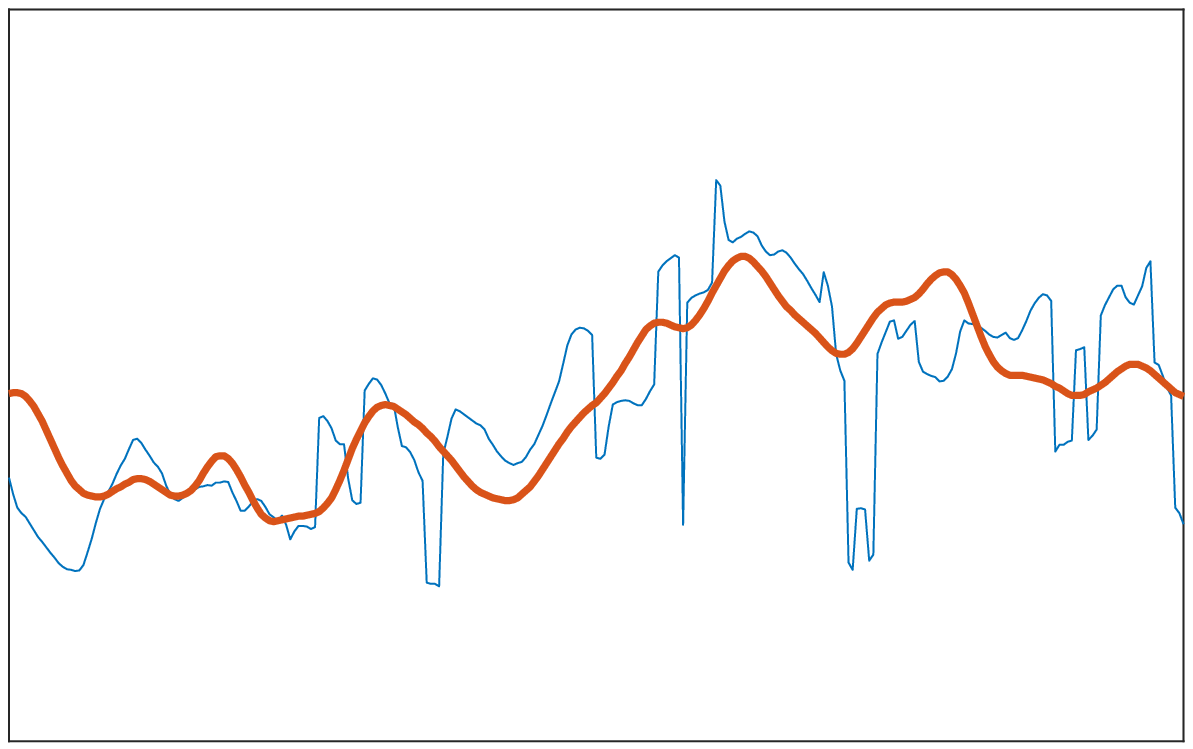}}
	\subfigure[$4$th enhanced.]{\includegraphics[width=1.15in]{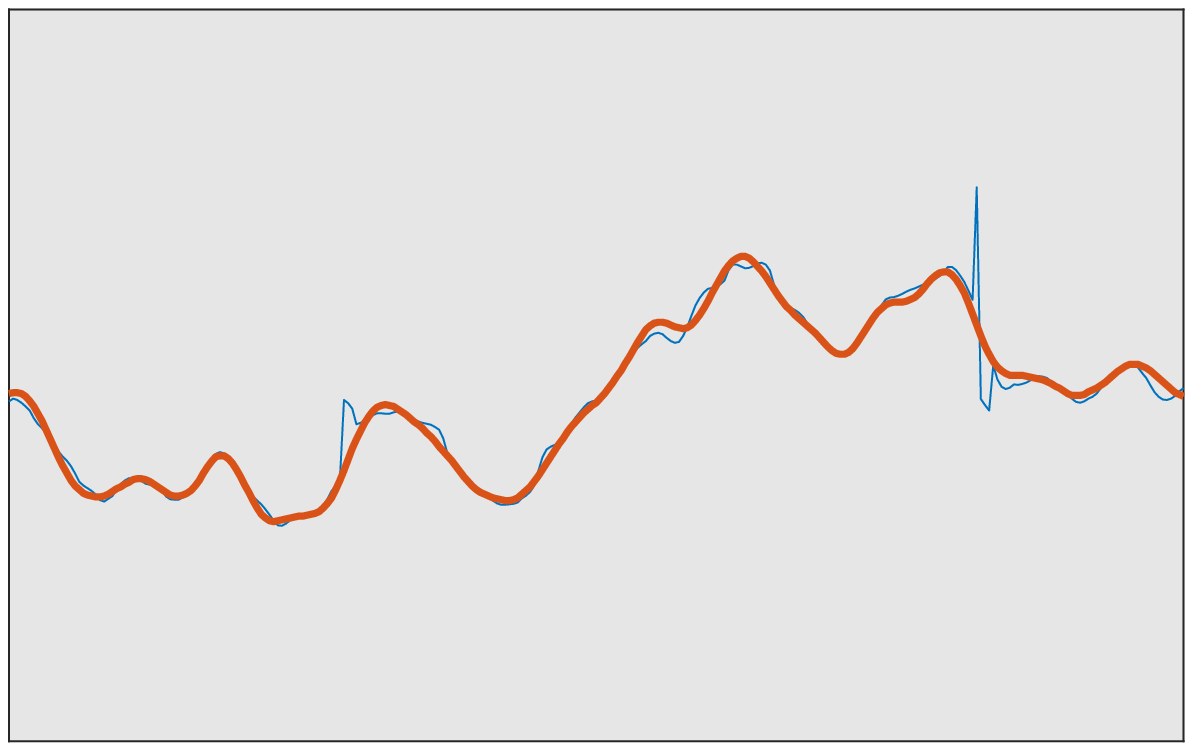}}
	\subfigure[$5$th enhanced.]{\includegraphics[width=1.15in]{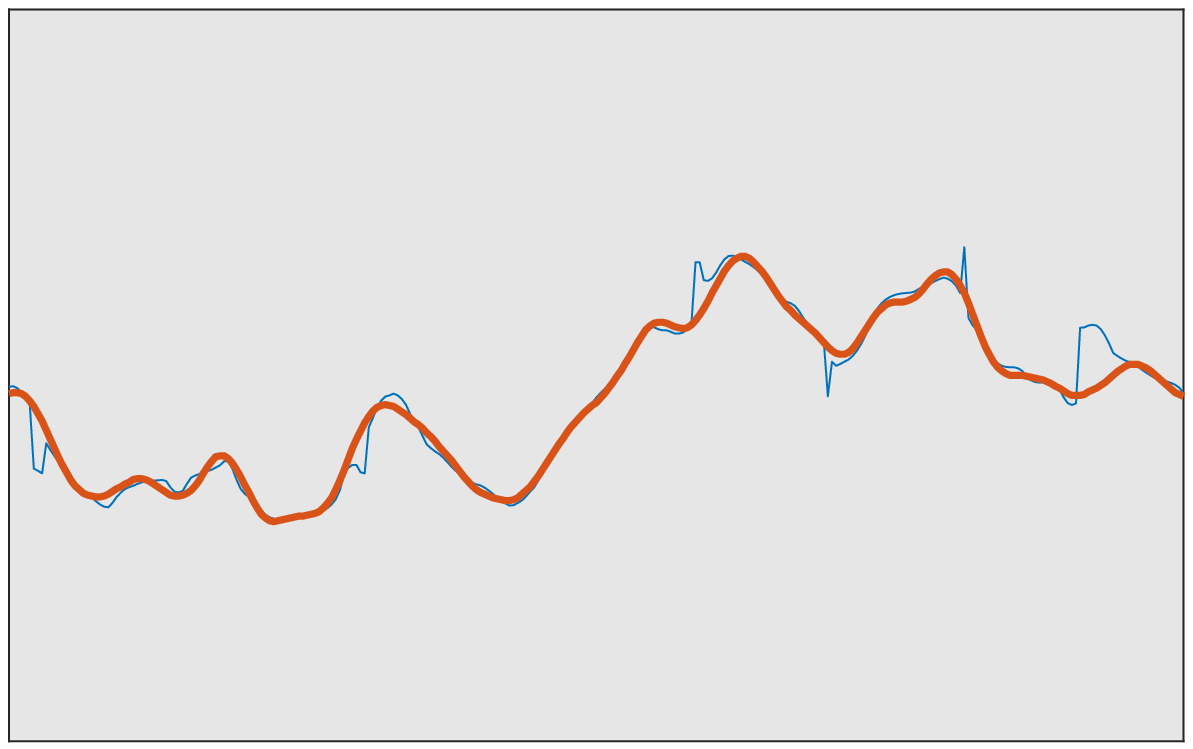}}
	\subfigure[$6$th enhanced.]{\includegraphics[width=1.15in]{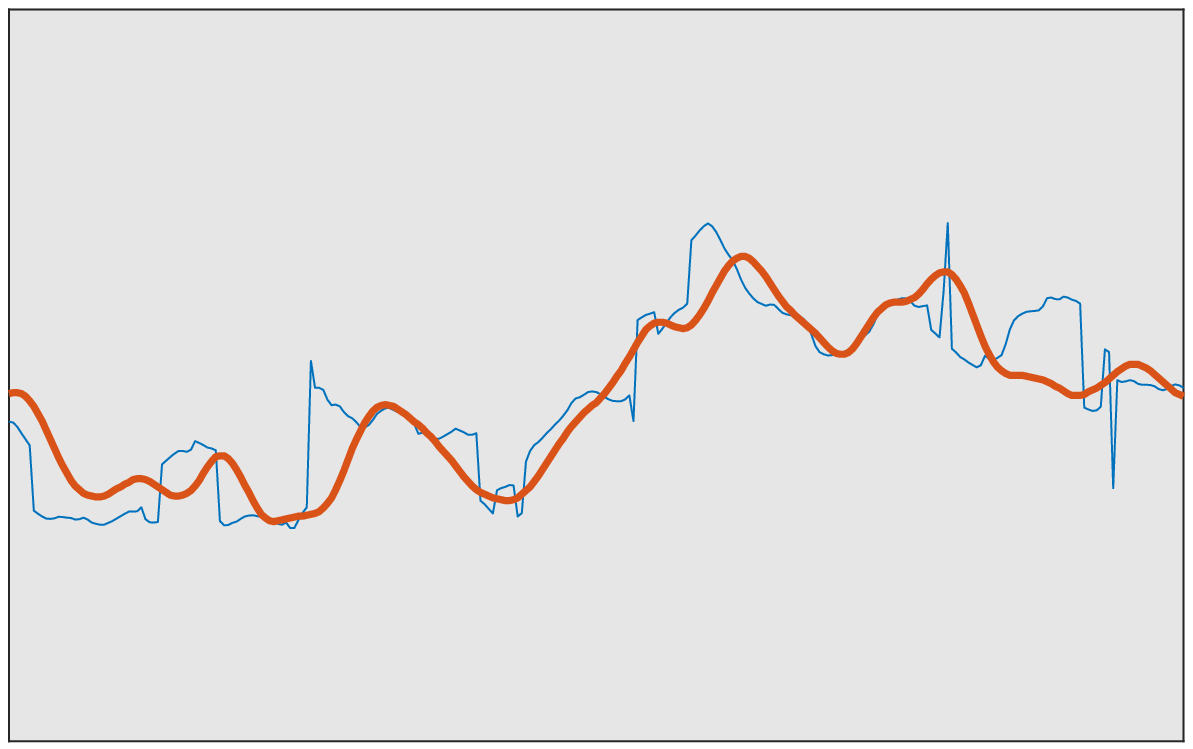}}
	\subfigure[$7$th enhanced.]{\includegraphics[width=1.15in]{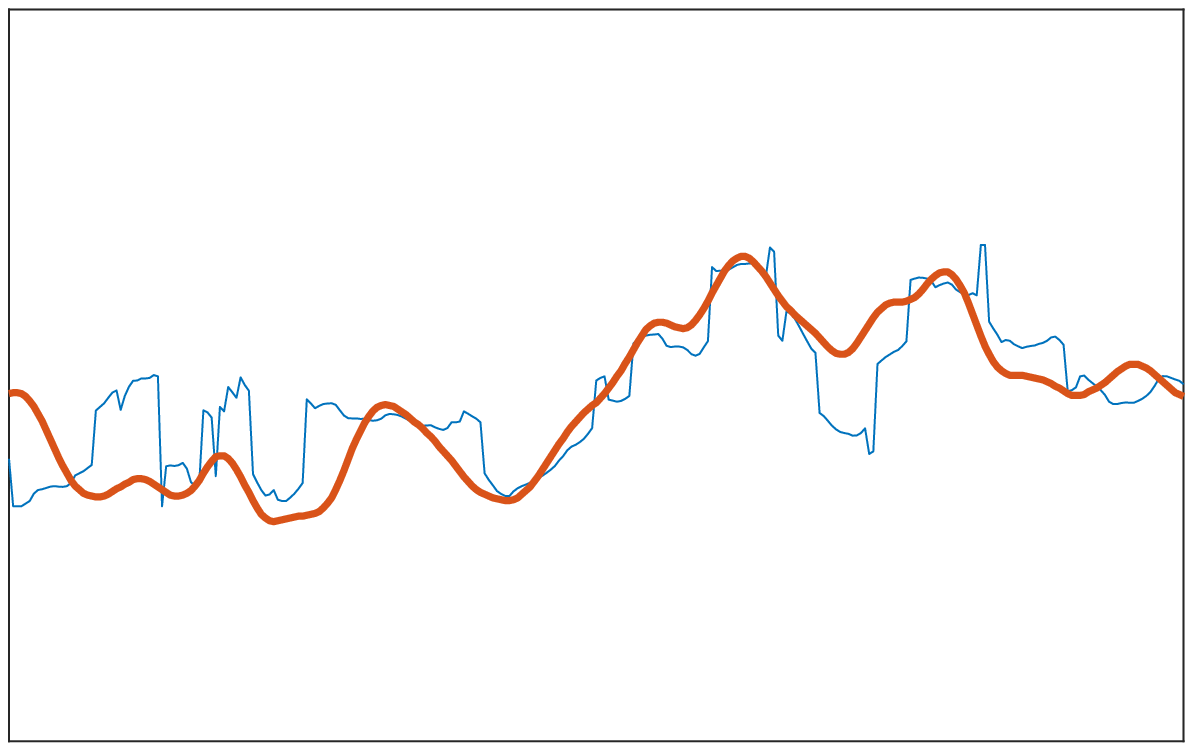}}\\
	\subfigure[$f_\text{single}{[l]}$, $41\times 10^{-5}$.]{\includegraphics[width=1.39in]{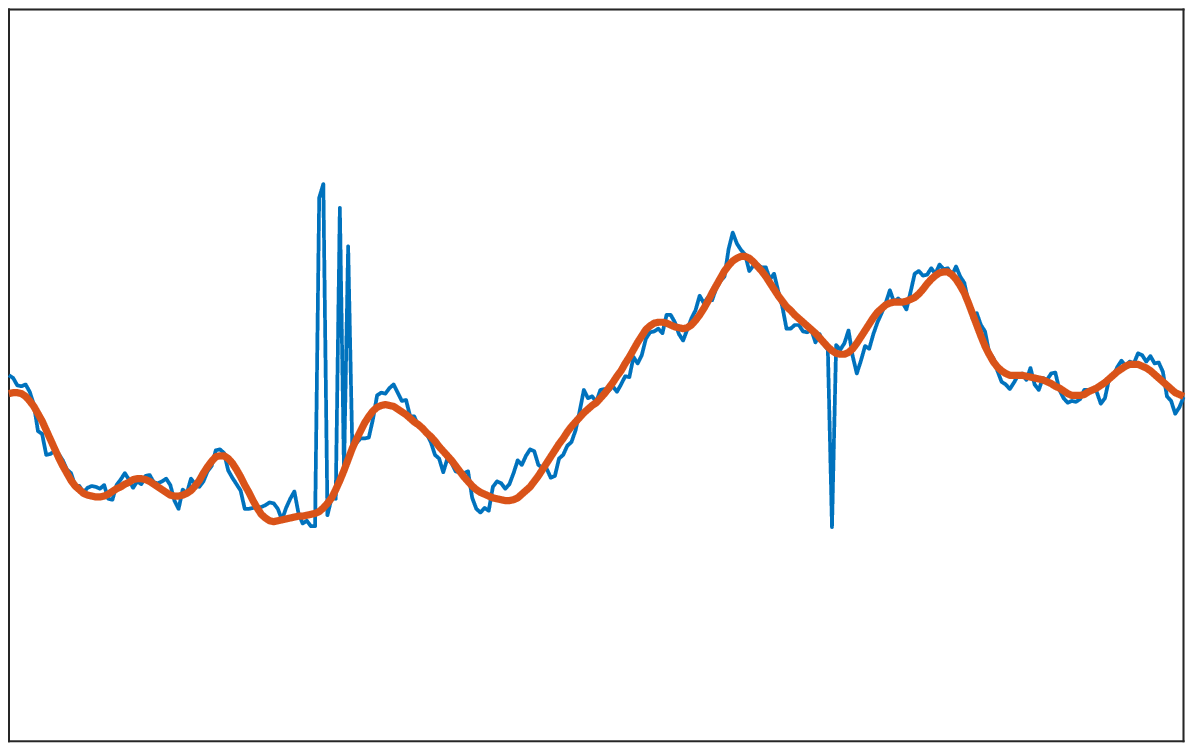}}
	\subfigure[$f_\text{E-single}{[l]}$, $2.9\times 10^{-5}$.]{\includegraphics[width=1.39in]{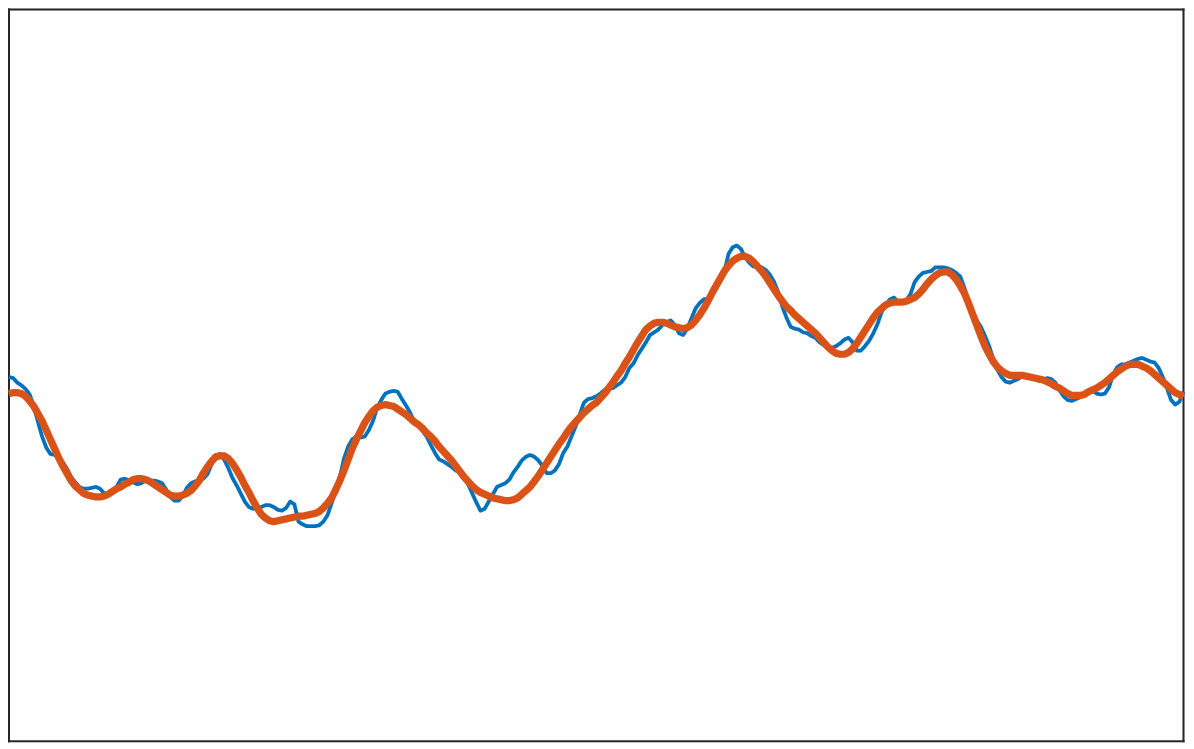}}
	\subfigure[$f_\text{MLE}{[l]}$, $2.5\times 10^{-5}$.]{\includegraphics[width=1.39in]{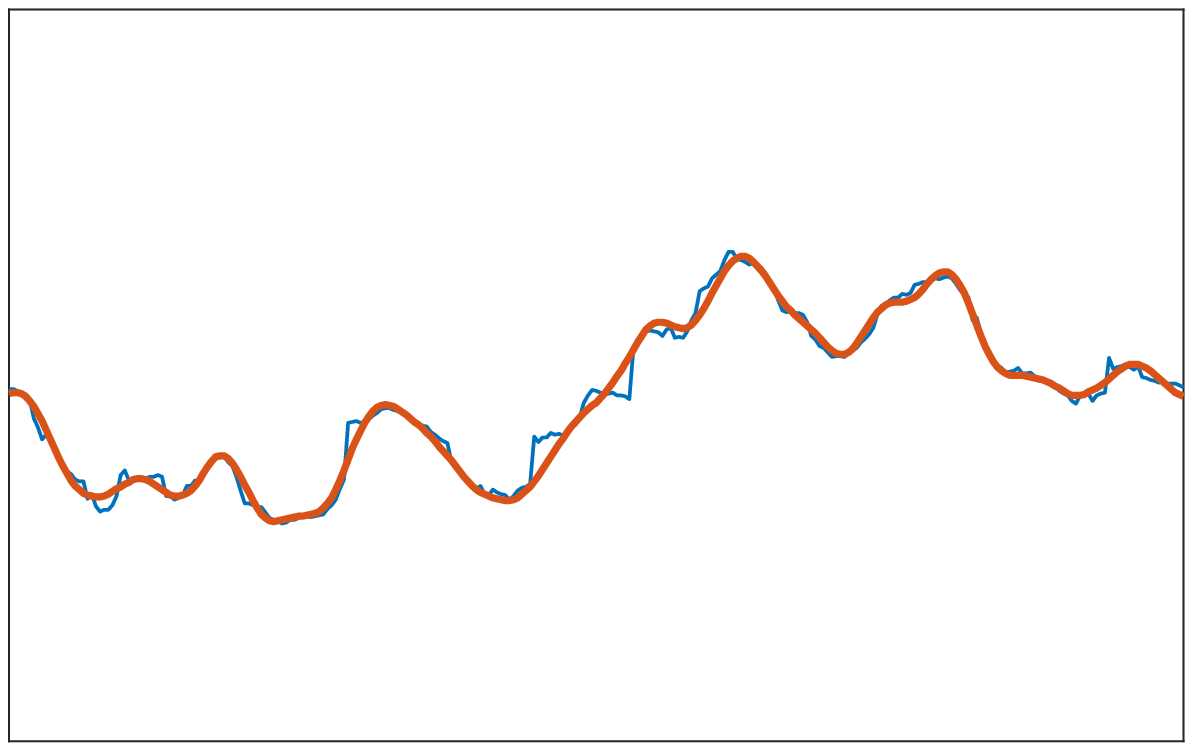}}
	\subfigure[$f_\text{WMLE}{[l]}$, $23\times 10^{-5}$.]{\includegraphics[width=1.39in]{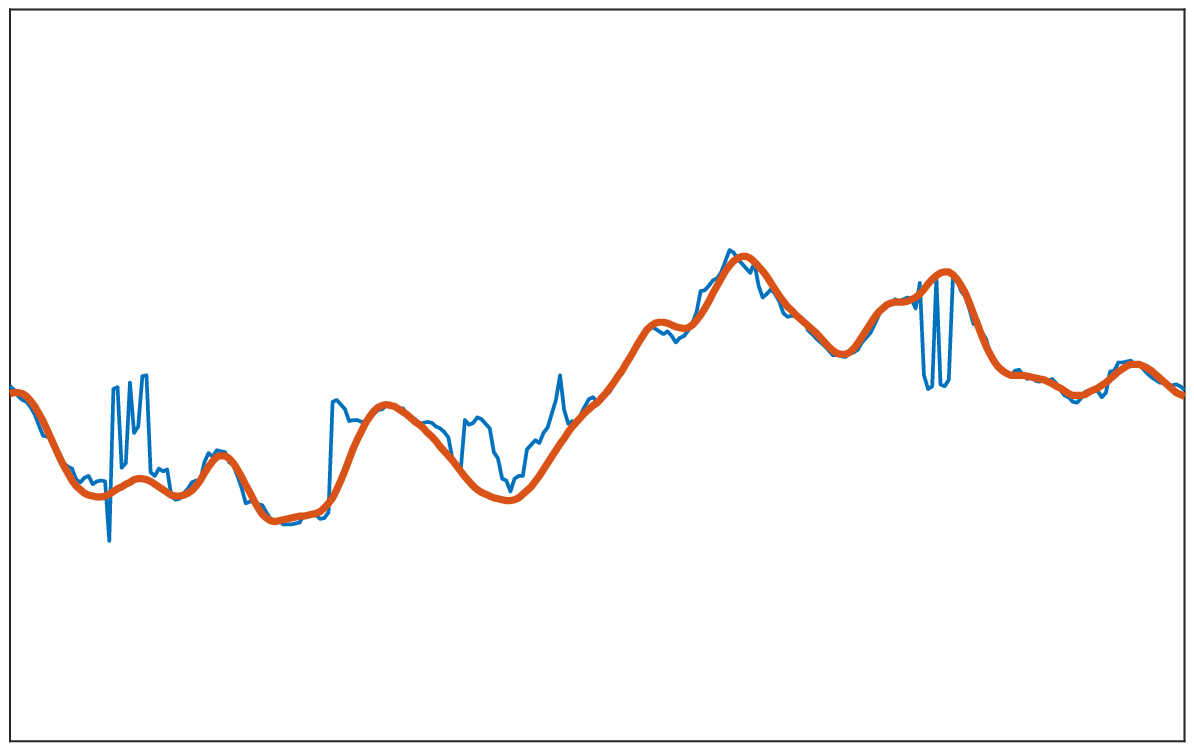}}
	\subfigure[$f_\text{E-MLE}{[l]}$, $2.3\times 10^{-5}$.]{\includegraphics[width=1.39in]{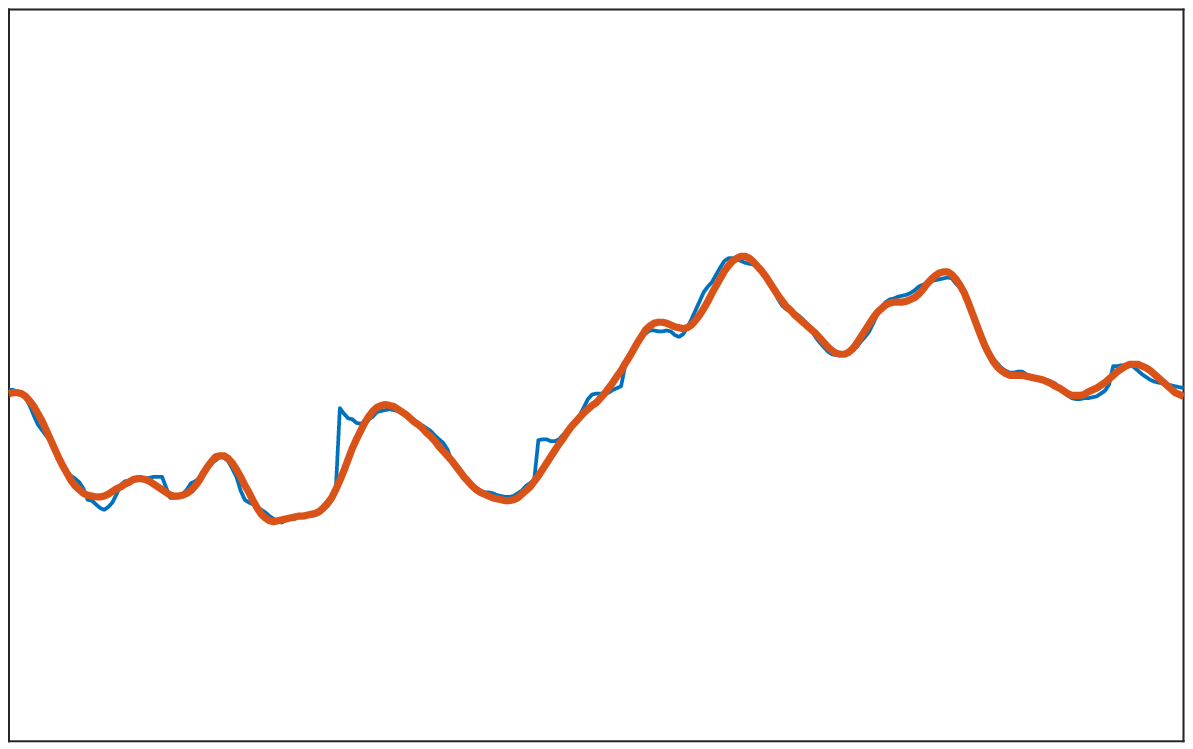}}\\
	\subfigure[$f_\text{E-WMLE}{[l]}$, $35\times 10^{-5}$.]{\includegraphics[width=1.39in]{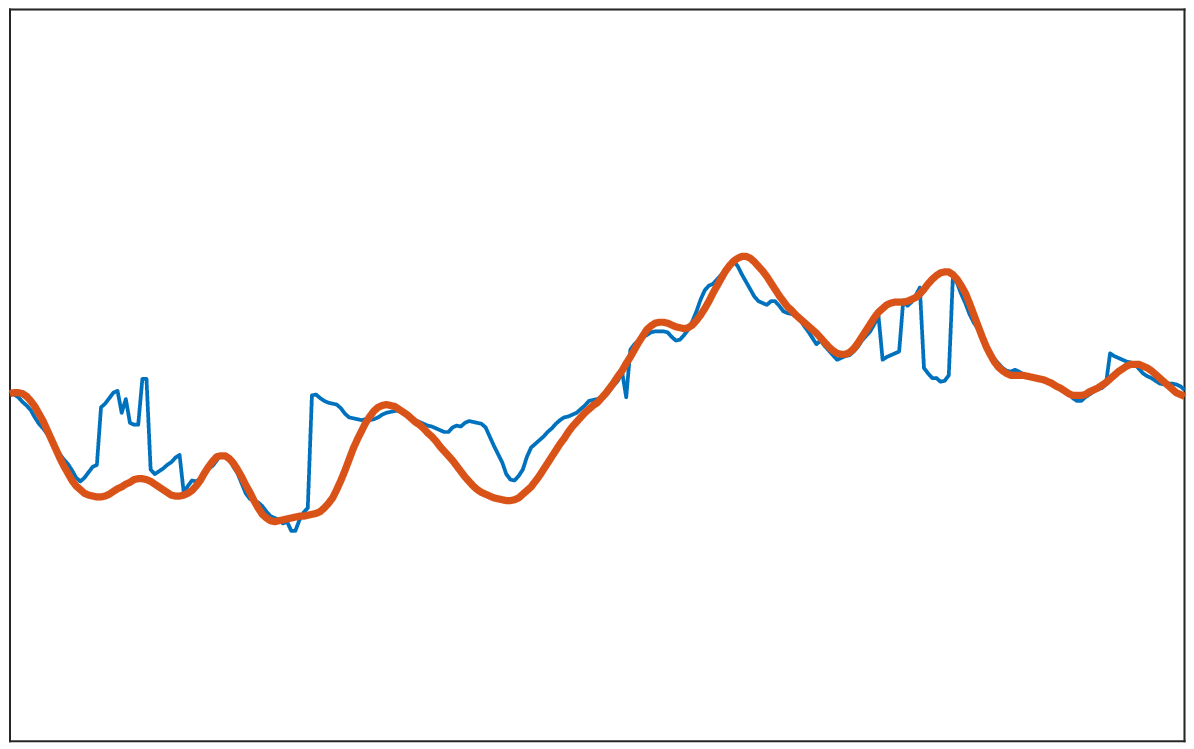}}
	\subfigure[$f_\text{S-MLE}{[l]}$, $1.1\times 10^{-5}$.]{\includegraphics[width=1.39in]{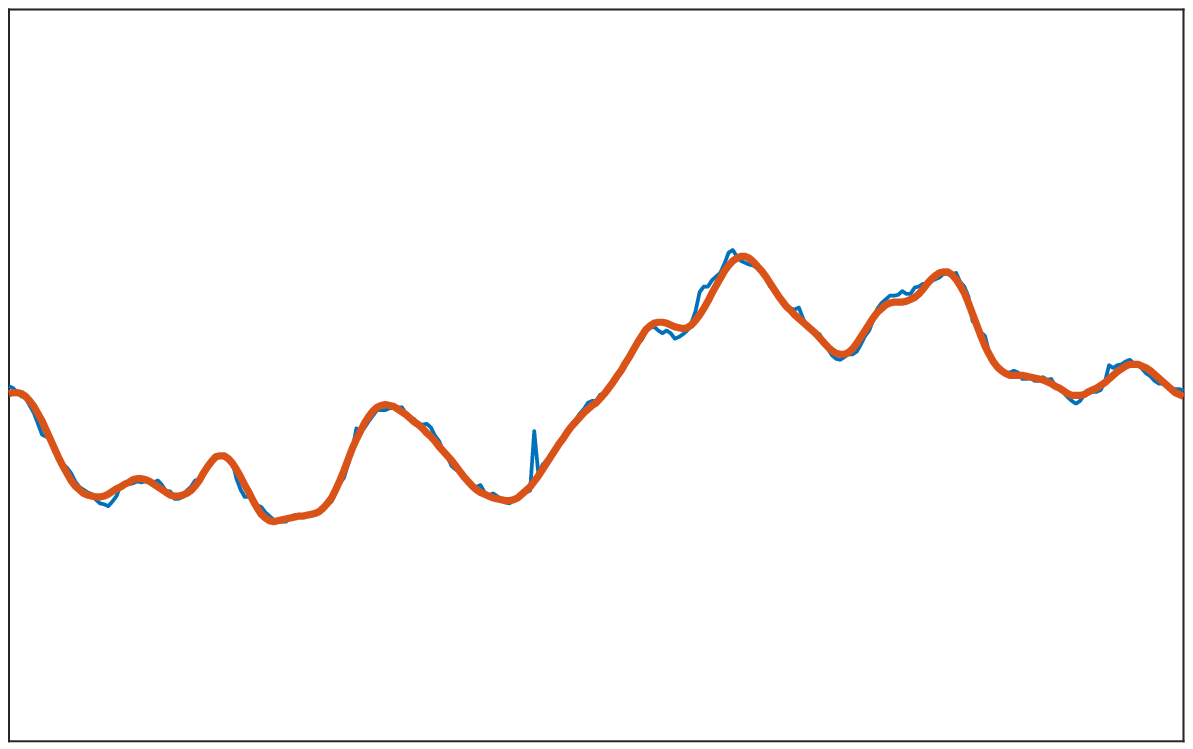}}
	\subfigure[$f_\text{S-WMLE}{[l]}$, $3.4\times 10^{-5}$.]{\includegraphics[width=1.39in]{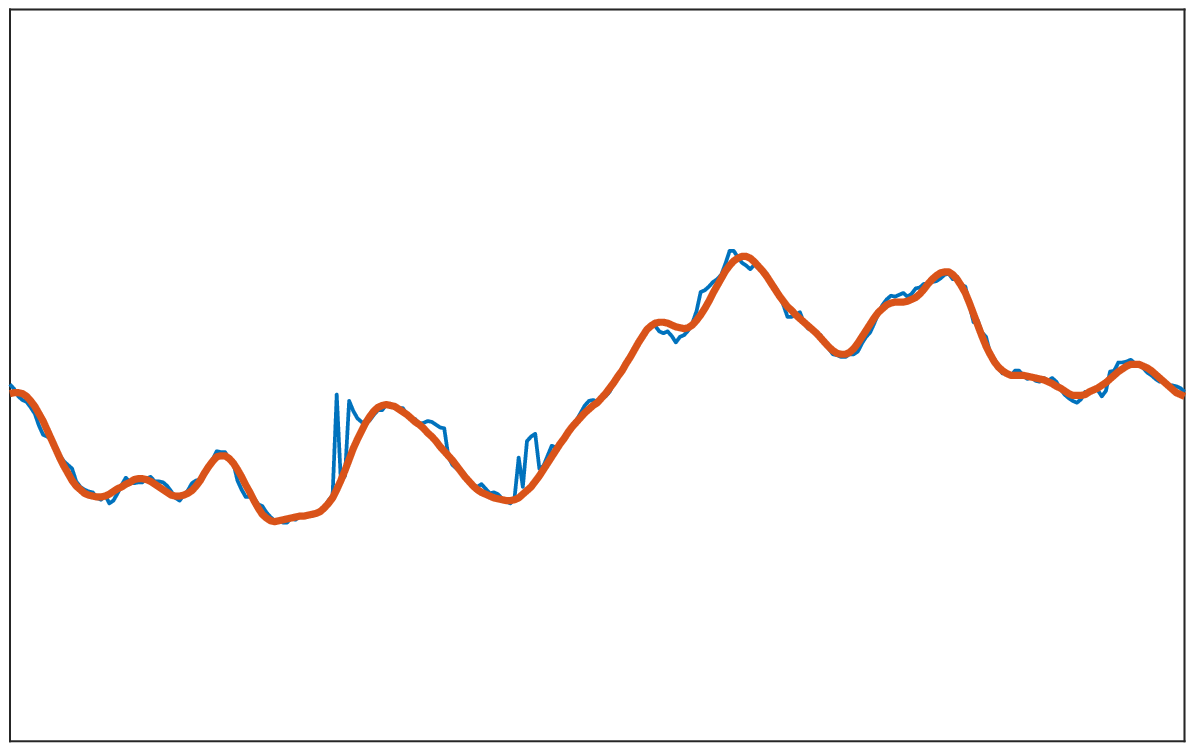}}
	\subfigure[$f_\text{P-MLE}{[l]}$, $1.1\times 10^{-5}$.]{\includegraphics[width=1.39in]{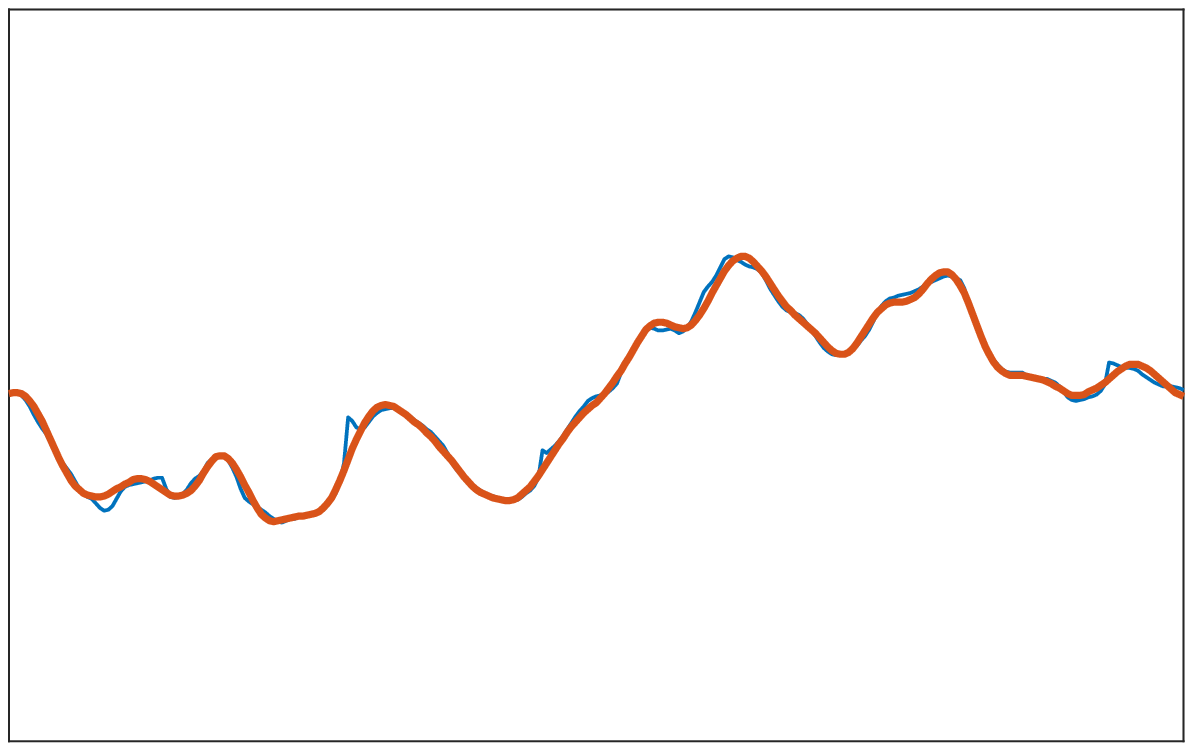}}
	\subfigure[$f_\text{P-WMLE}{[l]}$, $16\times 10^{-5}$.]{\includegraphics[width=1.39in]{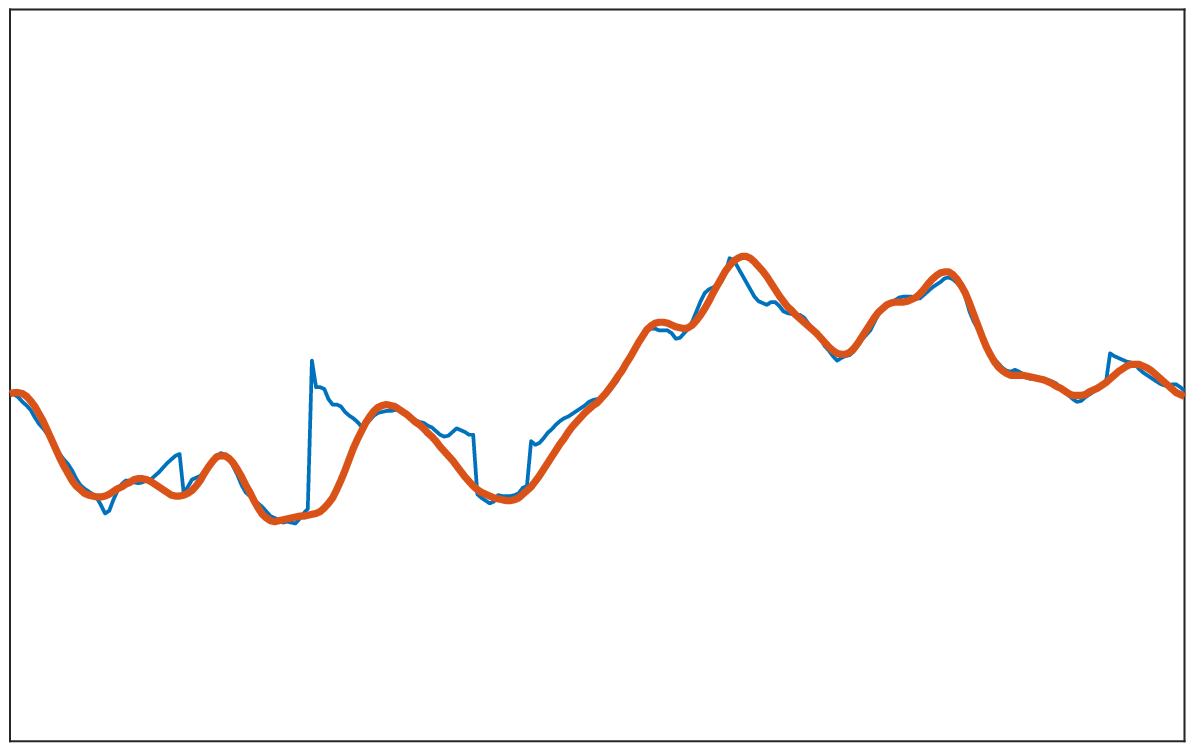}}
	\caption{A comparative demonstration of the single- and multi-tone ENF estimation schemes using a $5$-min synthetic signal, where $\operatorname{SNR}=-20$ dB, $\eta=0.8$, $\mathcal{M}=\{2\}$ for single-tone schemes, and $\mathcal{M}=\{2,3,4,5,6,7\}$ for multi-tone ones. All frequencies are normalized to the $2$nd harmonic band. For all subfigures, x-axis: sample index; y-axis: frequency (Hz); blue curve: estimated; red curve: ground truth. First row: individually evaluated harmonic components of the bandpass filtered synthetic signal $x[n]$. Second row: individually evaluated harmonic components of $x_\text{E}[n]$, i.e., the signal after the HRFA. Last two rows: the ENF estimation results (sub-captions present scheme notation and the corresponding MSE values) of the $10$ schemes summarized in Table \ref{estimator_summary}. The indices of corrupted harmonic components are $\{3,6,7\}$. Before the HRFA, the GHSA correctly yielded $\mathcal{C}=\{v_1,v_3,v_4\} \Rightarrow \Omega=\{2,4,5\}$ (first row gray ones), while after the HRFA, the GHSA selected an extra enhanced component, $\mathcal{C}=\{v_1,v_3,v_4,v_5\}\Rightarrow \Omega=\{2,4,5,6\}$ (second row gray ones).}
	\label{synthetic_demo}
\end{figure*}

\section{Performance Evaluation}
To evaluate the performances of the proposed algorithms and the overall framework, a series of schemes are implemented for comparison using both synthetic and real-world recordings, while all frequency estimates are based on STFT and are scaled to the $2$nd harmonic frequency band. Based on the choices of single- or multi-tone model, enhancement module, harmonic selection module, and the frequency estimator, the implemented schemes are listed in Table \ref{estimator_summary} and divided into $3$ groups. Group I consists of the state-of-the-art single- and multi-tone schemes, while the schemes in Group II are used to evaluate the performance of either HRFA or GHSA without applying the other. Group III contains the proposed framework employing both the HRFA and GHSA, with the use of both MLE and WMLE.

\emph{Global Parameter Setting:} The following parameters are set as constant values in this section unless otherwise mentioned. Considering general computational complexity and the fact that higher harmonic frequency bands are more likely to overlap with frequency bands of voice and audio content, we set $f_\text{S}=800$ Hz and thus $\mathcal{M}=\{2,3,4,5,6,7\}$, in which the fundamental frequency component is not considered as discussed in Section II-A. For the HRFA, according to the current reported results of the RFA \cite{Own_Singletone_Enhancement}, we set $I=2$, $\tau=3000$, and $\alpha={{f_{\text{S}}}/(4\max \left| {x[n]} \right|)}$. For the frame-based processing in the MLE and WMLE, we follow the conventional settings for ENF analysis, i.e.,  $\Delta=f_\text{S}$, $N_\text{F}=16f_\text{S}$, and the search regions for ENF harmonic estimation are set within $m \times [49.9,50.1]$ Hz. The signal and noise subbands for SNR estimation in the WMLE are set as $m\times [49.98, 50.02]$ Hz and $m\times \{[49, 49.98]\cup[50.02,51]\}$ Hz respectively. To ensure distinguishable ENF estimation differences among schemes, the frequency resolution of fast Fourier transform (FFT) is set to $1/4000$ Hz, i.e., a $(4000f_\text{S})$-point FFT is applied to each properly zero-padded frame. 

\subsection{A Demonstrative Example}
\begin{figure}[!t]
	\centering
	\includegraphics[width=3.35in]{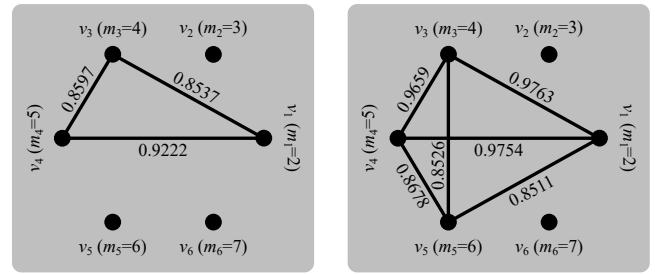}
	\caption{The corresponding graphs before (left) and after (right) the HRFA, generated using the synthetic signal in Fig. \ref{synthetic_demo}.}
	\label{graph_synthetic}
\end{figure}

We first provide an example using a synthetic signal to demonstrate the performances of the ENF estimation schemes, in which the ENF is generated using a first-order autoregressive (AR) model. Let $f_\mathcal{N}[n]\sim \mathcal{N}(0,1)$, $n\in\{0,1,\ldots,N-1\}$, then the zero-mean synthetic fundamental frequency is given by \cite{App_Gaussian_Model}
\begin{equation}
f_\text{GT}[n]=0.99f_\text{GT}[n-1]+f_\mathcal{N}[n],
\end{equation}
which is then scaled to have a variance of $4.5\times10^{-4}$ and shifted to have a mean of $50$, according to Central China Grid. After that, the integer multiples $\mathcal{M}=\{2,3,4,5,6,7\}$ of the fundamental component are substituted into (\ref{original_IF_model}). To simulate practical situations, perturbations are added to a randomly selected subset of $\mathcal{M}$ ($\{3,6,7\}$ in this case). The corresponding results are shown in Figs. \ref{synthetic_demo} and \ref{graph_synthetic}.

The results observed from the two figures are summarized as follows. \textbf{i)} Comparing the first and second rows of Fig. \ref{synthetic_demo}, all the harmonic components are effectively enhanced with the use of the HRFA, but for the $\{3,6,7\}$th corrupted components the IF details are not fully recovered. \textbf{ii)} After the HRFA, $4$ instead of originally $3$ harmonic components are selected by the GHSA, as shown from the gray subfigures of Fig. \ref{synthetic_demo}, whose graphical representation constructed from the correlation matrix $\mathbf{R}_\text{adj}$ is shown in Fig. \ref{graph_synthetic}. The enhanced harmonic components are seen to be more mutually correlated. \textbf{iii)} Figs. \ref{synthetic_demo} (a) and (g) are identical to Figs. \ref{synthetic_demo} (m) and (n), which are single-tone estimates without and with enhancement respectively. The effectiveness of the RFA \cite{Own_Singletone_Enhancement} in single-tone scenario is verified here. \textbf{iv)} In this synthetic example, the MLE-based schemes have generally yielded better results than the WMLE-based ones, and the reason lies in the inaccurate estimations of local SNR values. \textbf{v)} For the MLE-based schemes, the ones with the HRFA only in Fig. \ref{synthetic_demo} (q), with the GHSA only in Fig. \ref{synthetic_demo} (s), and with both the HRFA and GHSA in Fig. \ref{synthetic_demo} (u), all achieve lower mean squared error (MSE) values than directly applying the MLE in Fig. \ref{synthetic_demo} (o). \textbf{vi)} Comparing the best results in Figs. \ref{synthetic_demo} (s) and (u), we observe that the scheme with only harmonic selection where $\Omega=\{2,4,5\}$ achieves the same MSE value as the one with both enhancement and harmonic selection where $\Omega=\{2,4,5,6\}$. Meanwhile, comparing the results in Figs. \ref{synthetic_demo} (n) and (o), the single-tone estimator using the RFA achieves very competitive result against the conventional MLE, indicating that using an enhanced single tone is possible to achieve almost the same performance of using $6$ unenhanced harmonic components. This reveals an important trade-off when increasing the number of harmonic components for ENF estimation. It can be seen from Fig. \ref{synthetic_demo} (k) that the newly added $6$th component (after the HRFA) is still noisy, whose CC values with other components are all below $0.9$ as shown in Fig. \ref{graph_synthetic}. In fact, the noise in this component has compensated the performance improvement of adding this component. Therefore, it is verified that for a multi-tone ENF estimator, using more harmonic components may not always yield better results. This will be more sufficiently verified using real-world examples in Section IV-C. 

\begin{figure}[!t]
	\centering
	\includegraphics[width=3.35in]{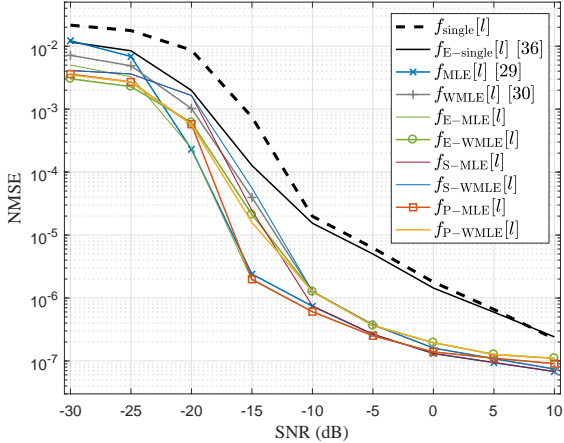}
	\caption{Performances of the ENF estimation schemes in terms of NMSE versus SNR using a $3$-min synthetic recording. Results are obtained via $100$ Monte Carlo experiments. The experimental setting favors $f_\text{MLE}[l]$ \cite{Estimation_Harmonics} since the noise is WGN, having equal effects on each harmonic component.}
	\label{synthetic_MSE_SNR_ALL}
\end{figure}

\begin{figure}[!t]
	\centering
	\includegraphics[width=3.35in]{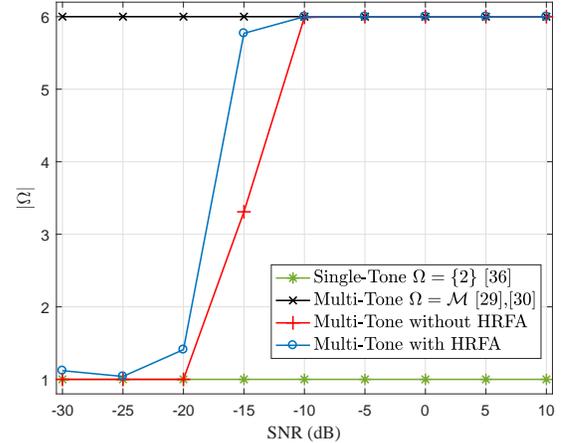}
	\caption{Comparison of the numbers of selected harmonic components $|\Omega|$ versus SNR using a $3$-min synthetic recording for the ENF estimation schemes, where $\mathcal{M}=\{2,3,4,5,6,7\}$ and $\eta=0.8$. Results are obtained via $100$ Monte Carlo experiments.}
	\label{synthetic_Omega_SNRs}
\end{figure}

\begin{figure*}[!t]
	\centering
	\subfigure[$2$nd.]{\includegraphics[width=1.15in]{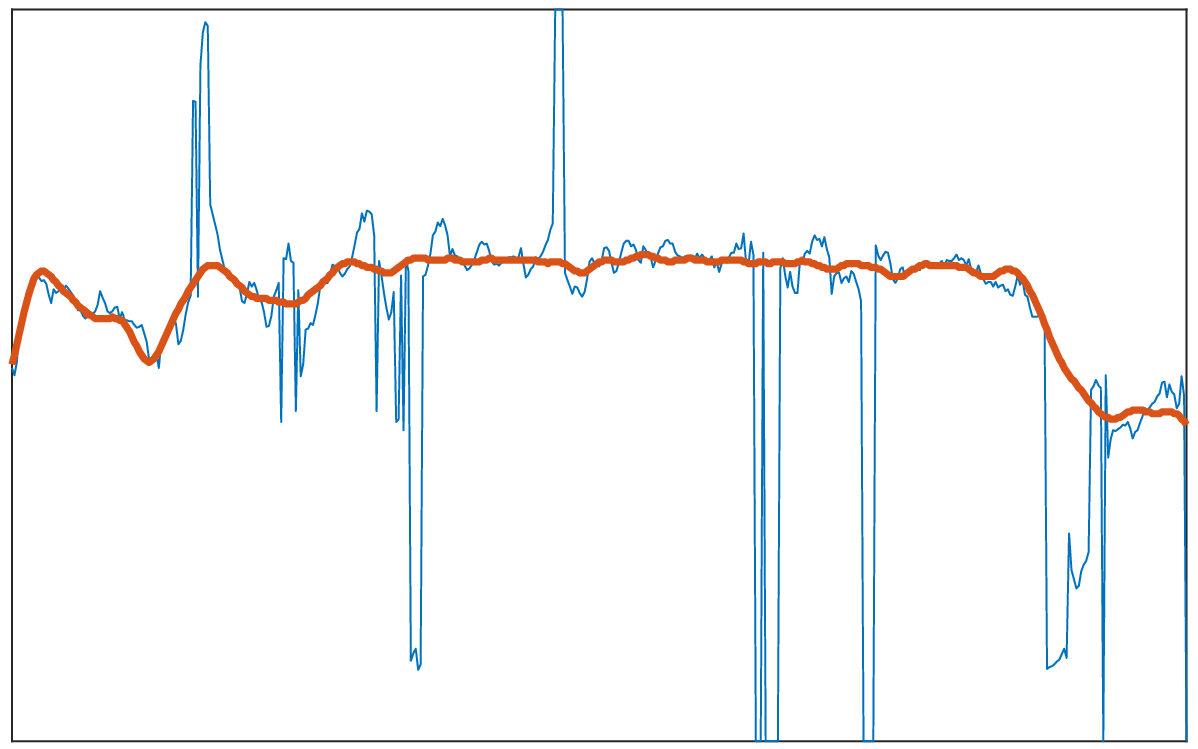}}
	\subfigure[$3$rd.]{\includegraphics[width=1.15in]{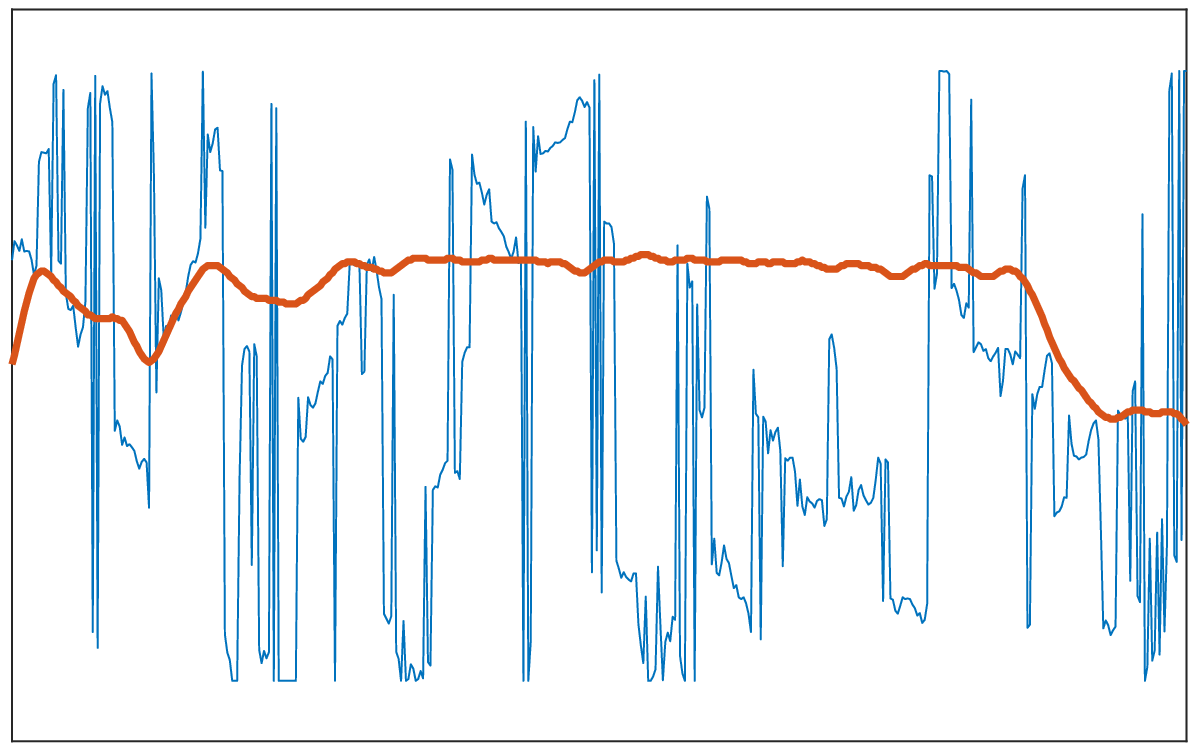}}
	\subfigure[$4$th.]{\includegraphics[width=1.15in]{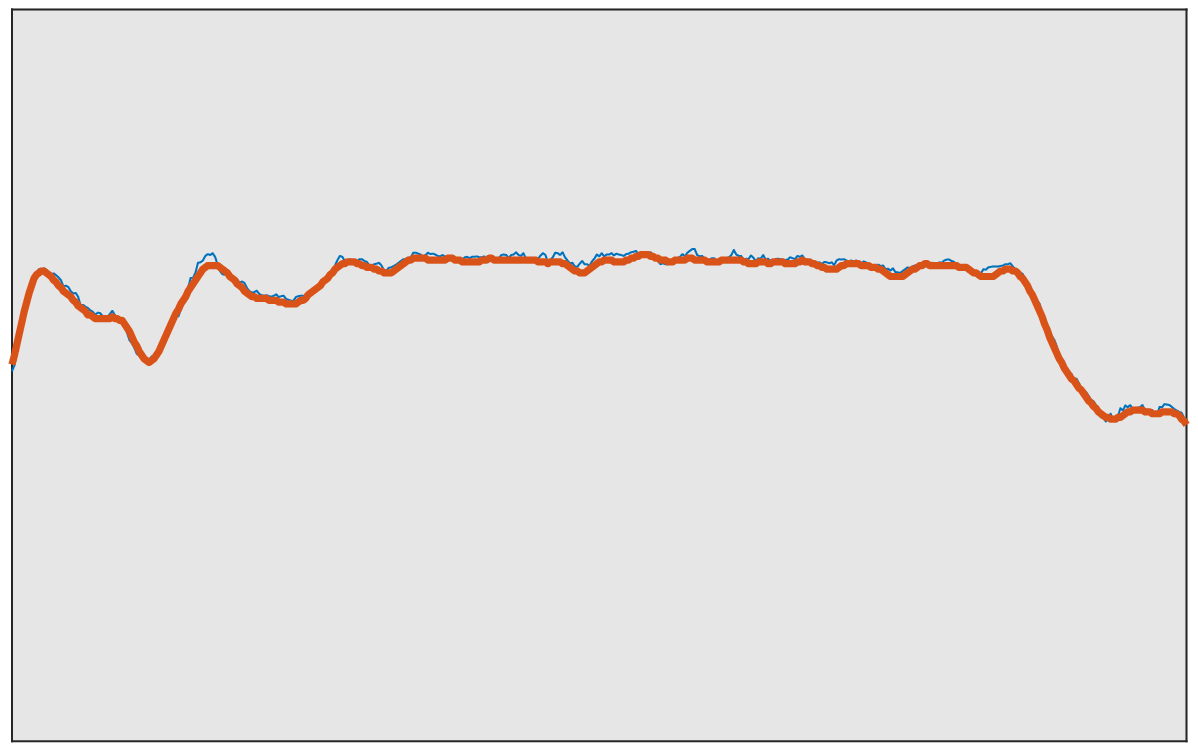}}
	\subfigure[$5$th.]{\includegraphics[width=1.15in]{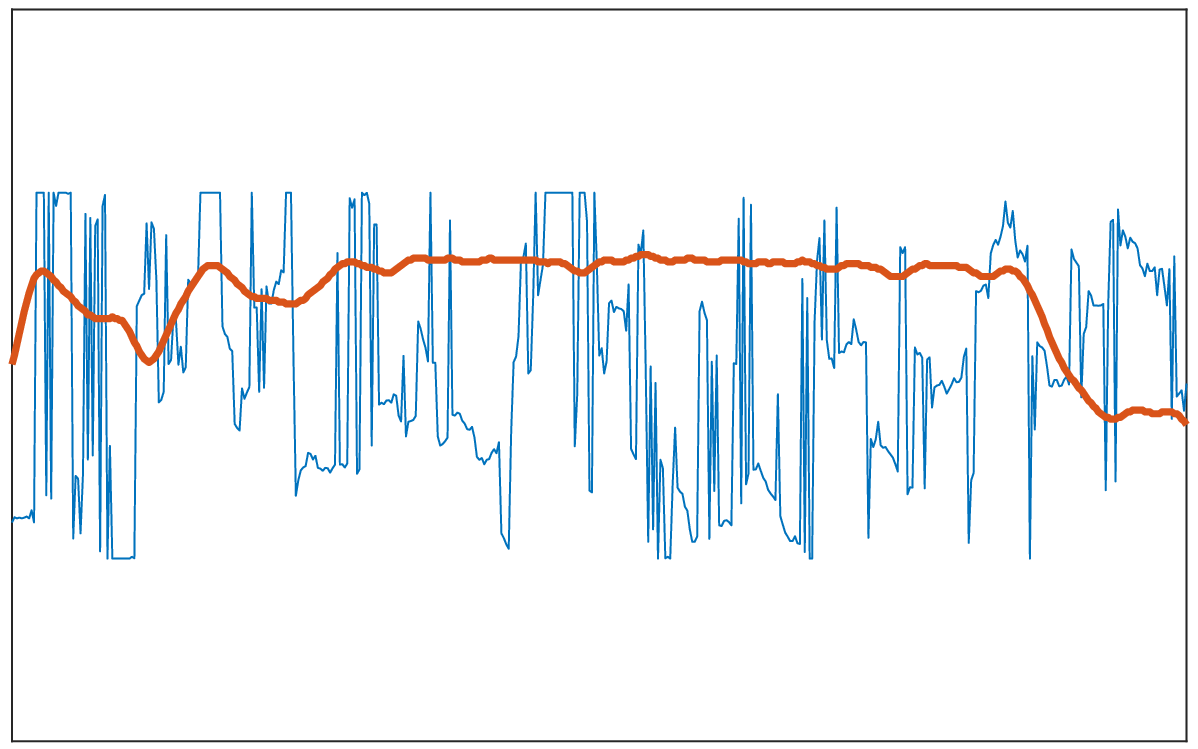}}
	\subfigure[$6$th.]{\includegraphics[width=1.15in]{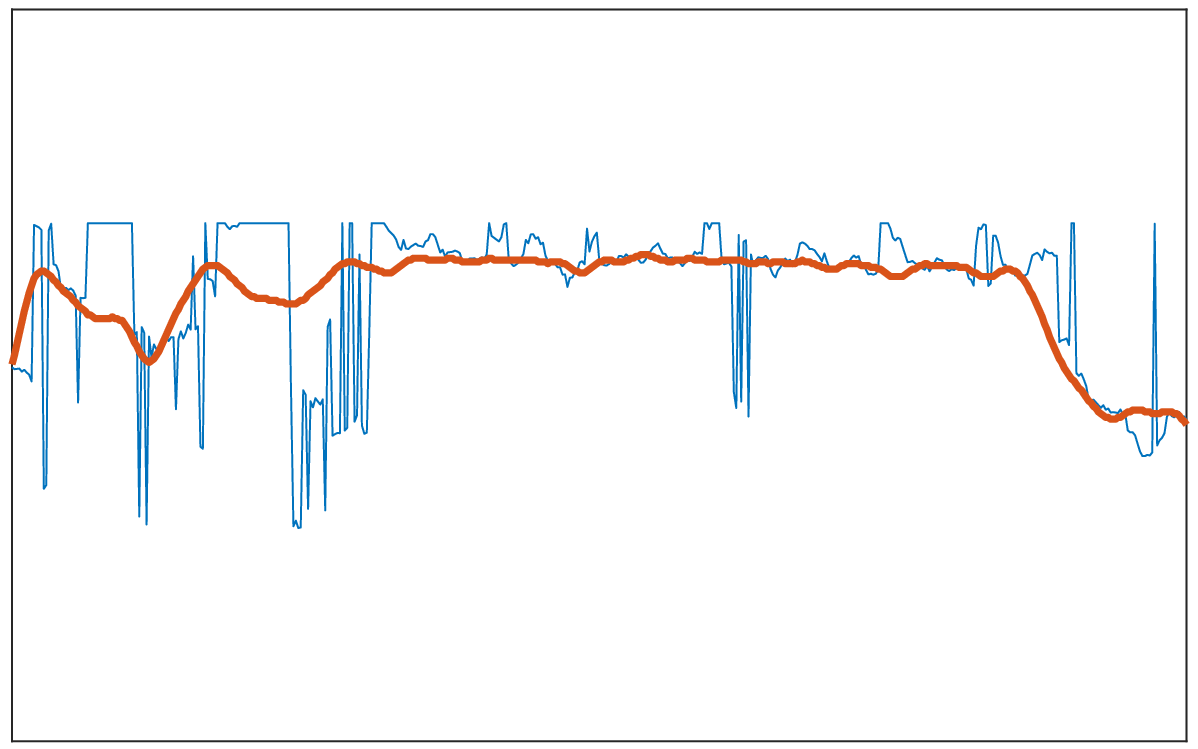}}
	\subfigure[$7$th.]{\includegraphics[width=1.15in]{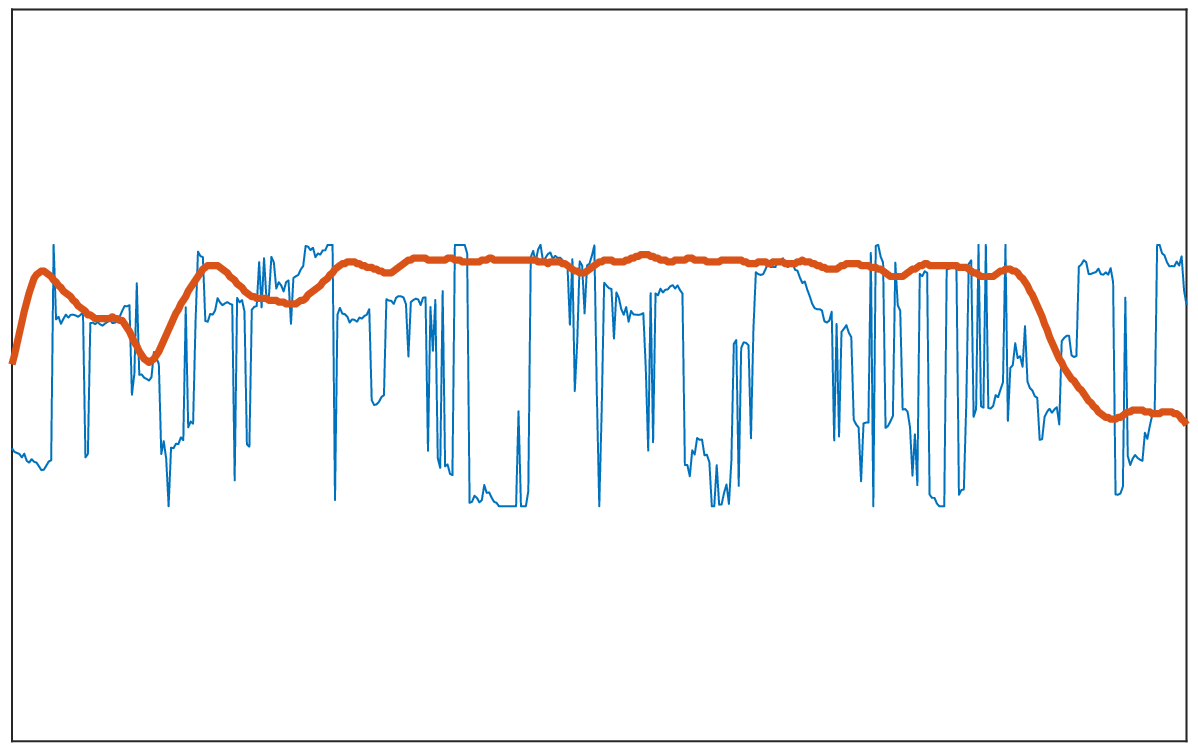}}\\
	\subfigure[$2$nd enhanced.]{\includegraphics[width=1.15in]{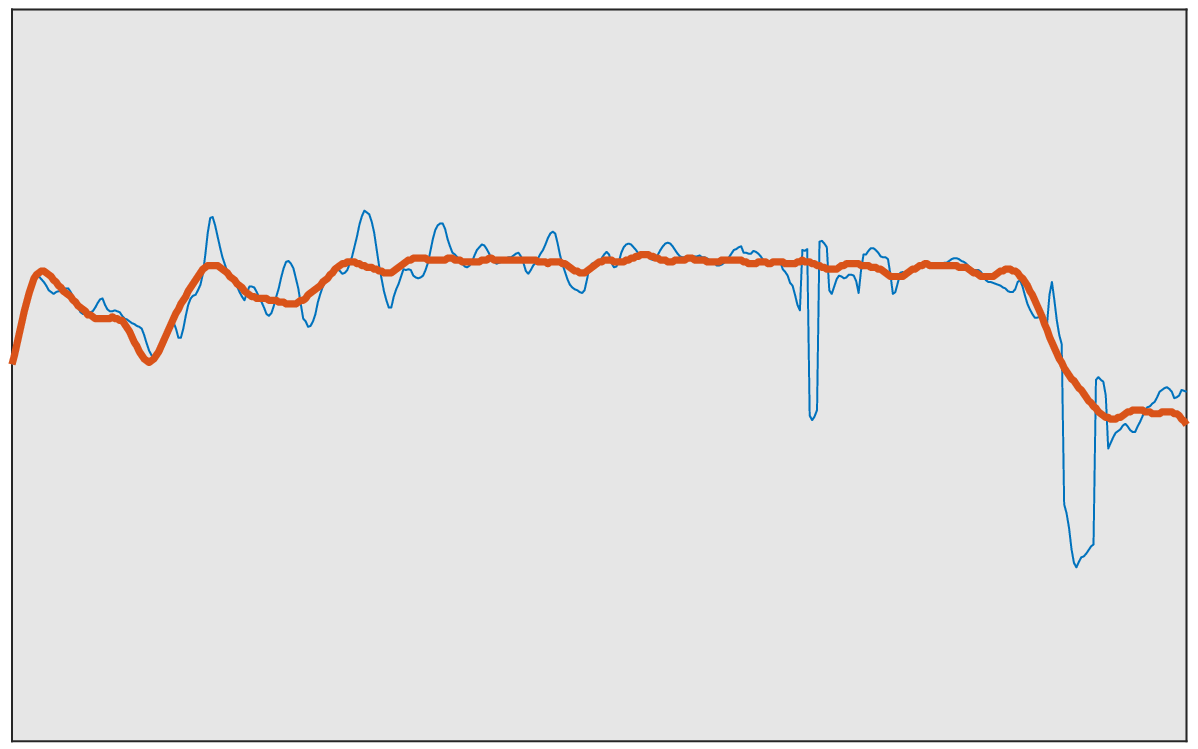}}
	\subfigure[$3$rd enhanced.]{\includegraphics[width=1.15in]{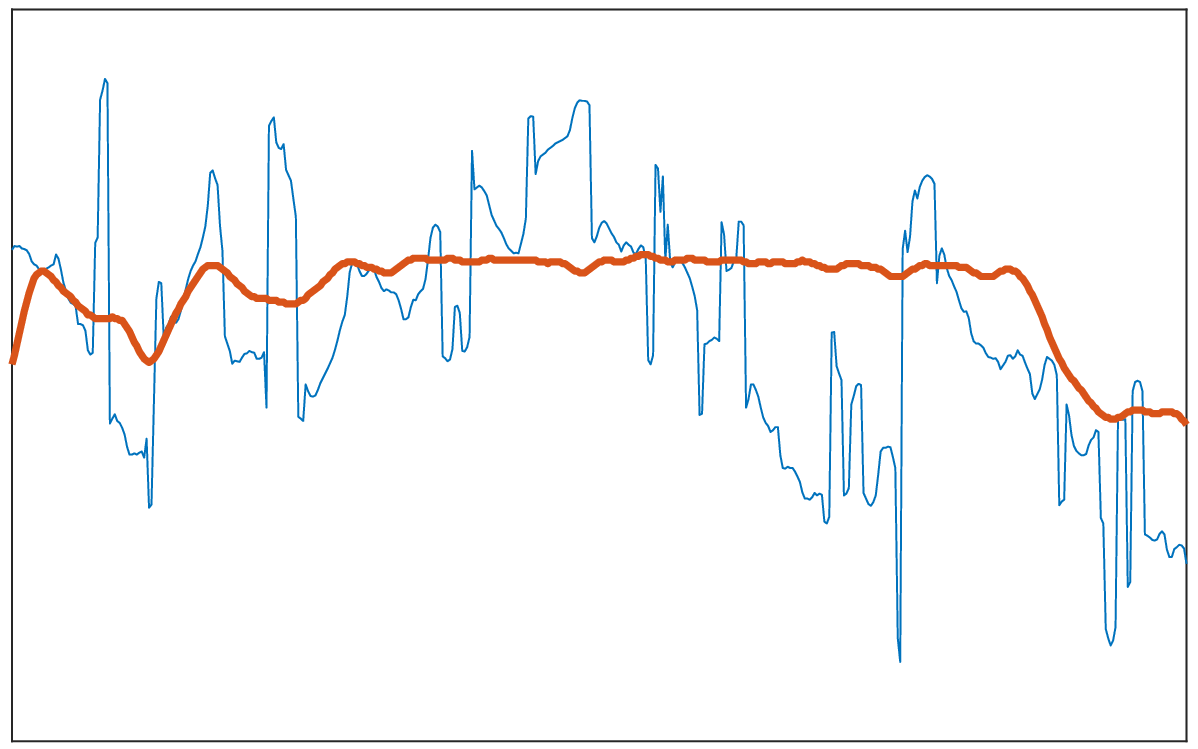}}
	\subfigure[$4$th enhanced.]{\includegraphics[width=1.15in]{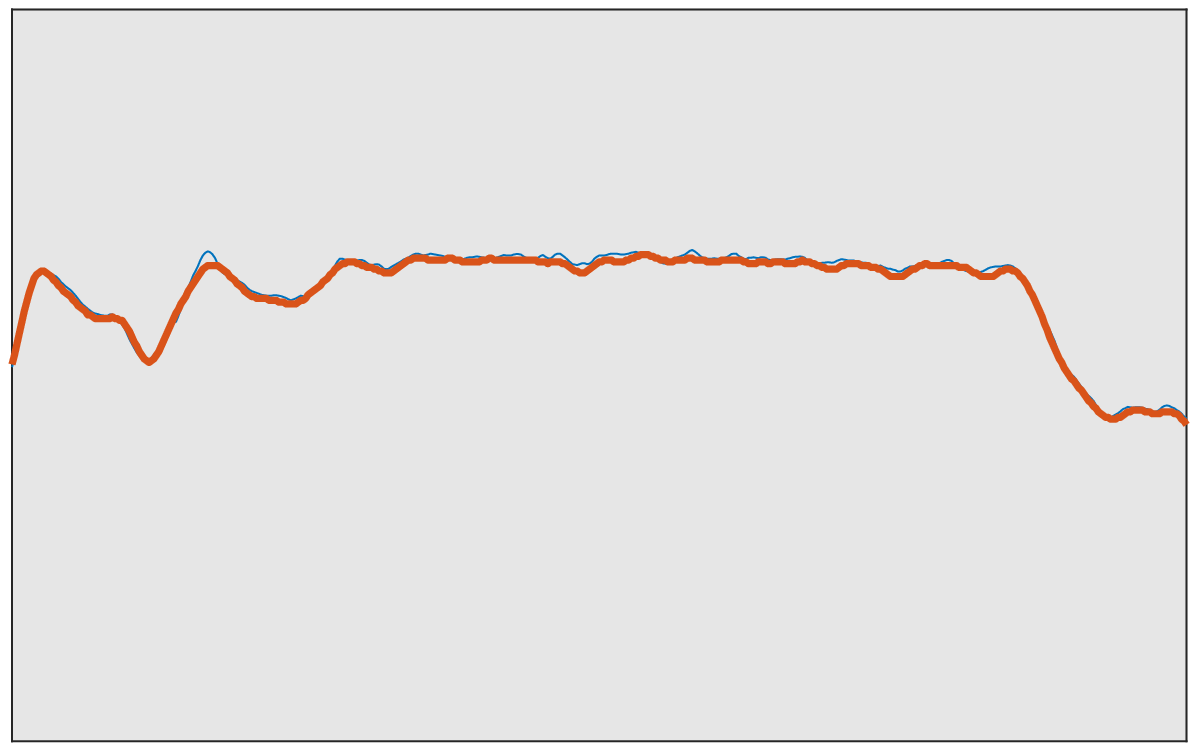}}
	\subfigure[$5$th enhanced.]{\includegraphics[width=1.15in]{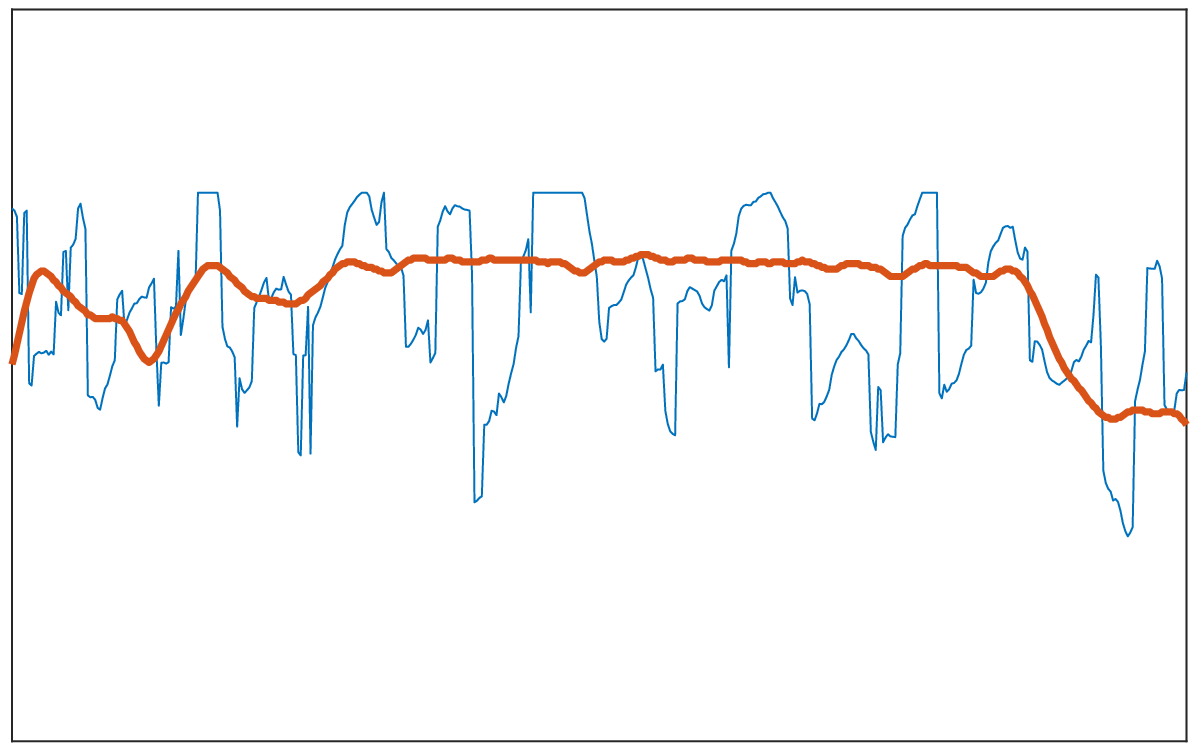}}
	\subfigure[$6$th enhanced.]{\includegraphics[width=1.15in]{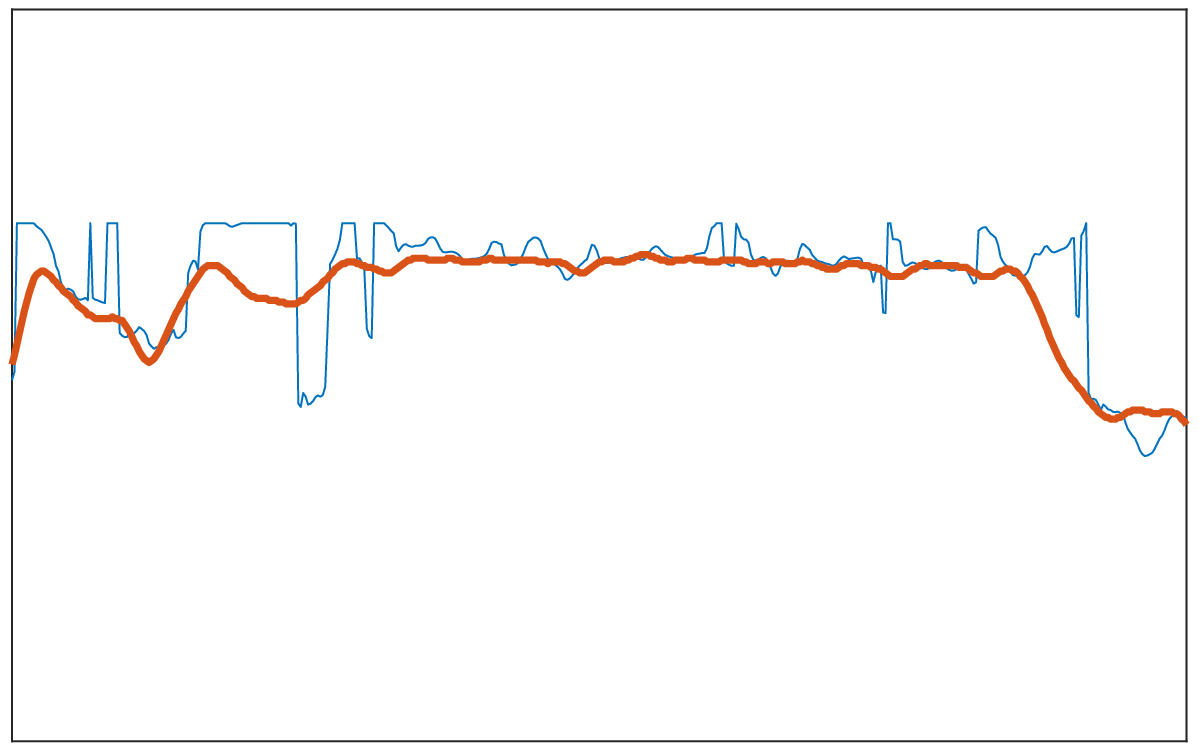}}
	\subfigure[$7$th enhanced.]{\includegraphics[width=1.15in]{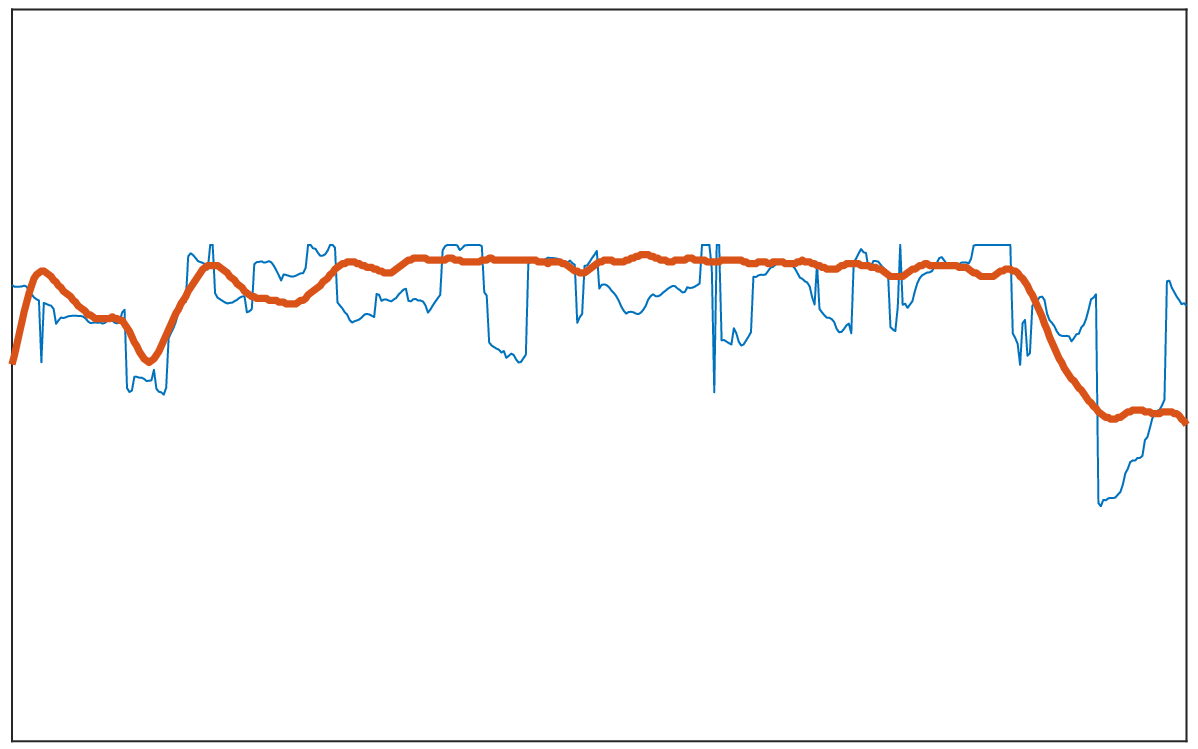}}\\
	\subfigure[$f_\text{single}{[l]}$, $380\times 10^{-5}$.]{\includegraphics[width=1.39in]{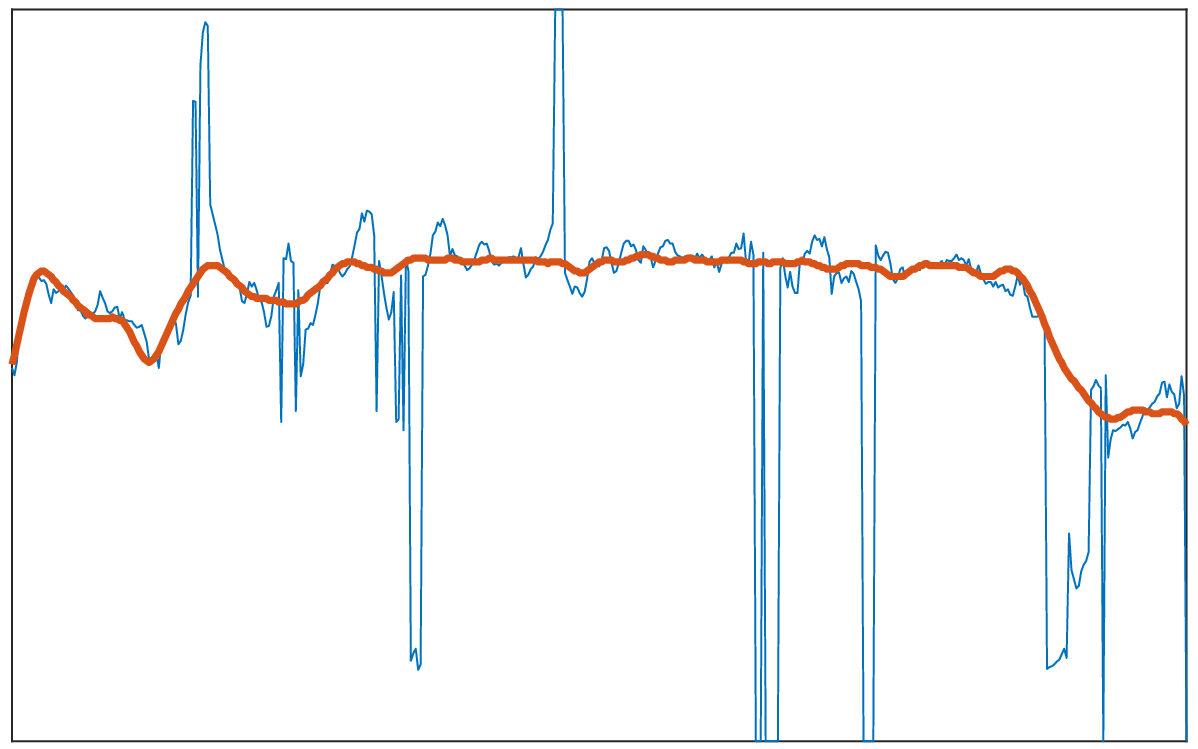}}
	\subfigure[$f_\text{E-single}{[l]}$, $34\times 10^{-5}$.]{\includegraphics[width=1.39in]{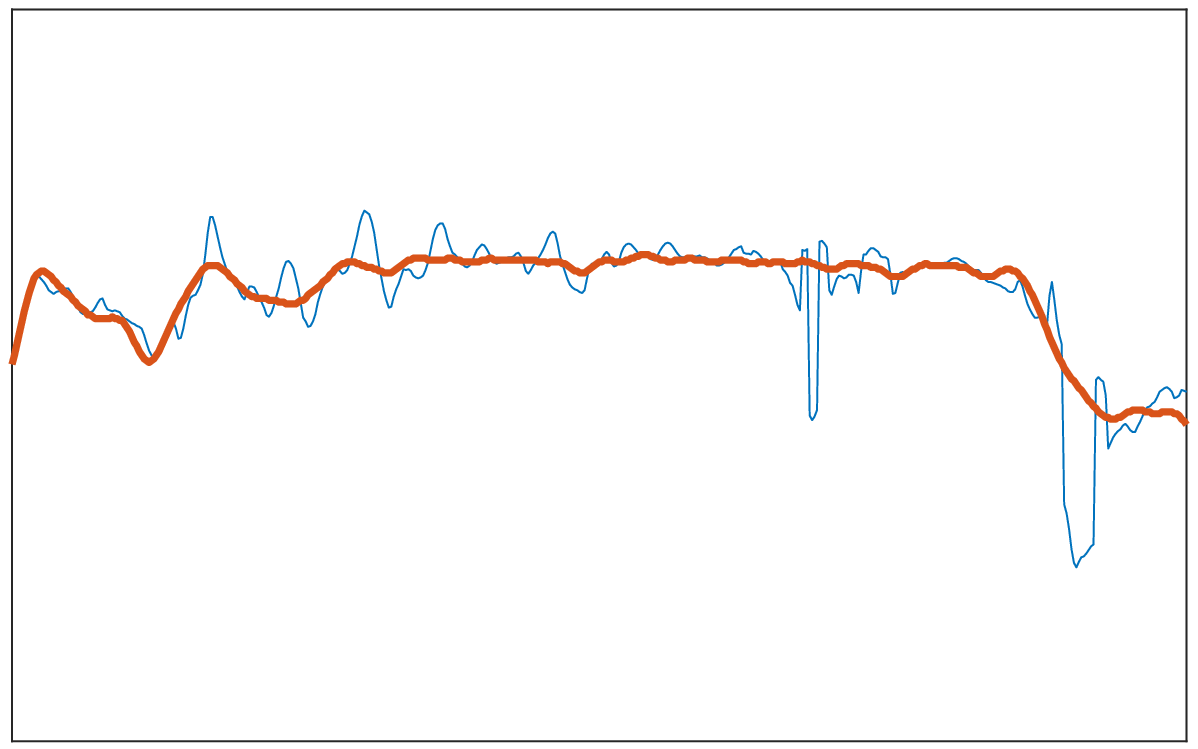}}
	\subfigure[$f_\text{MLE}{[l]}$, $28\times 10^{-5}$.]{\includegraphics[width=1.39in]{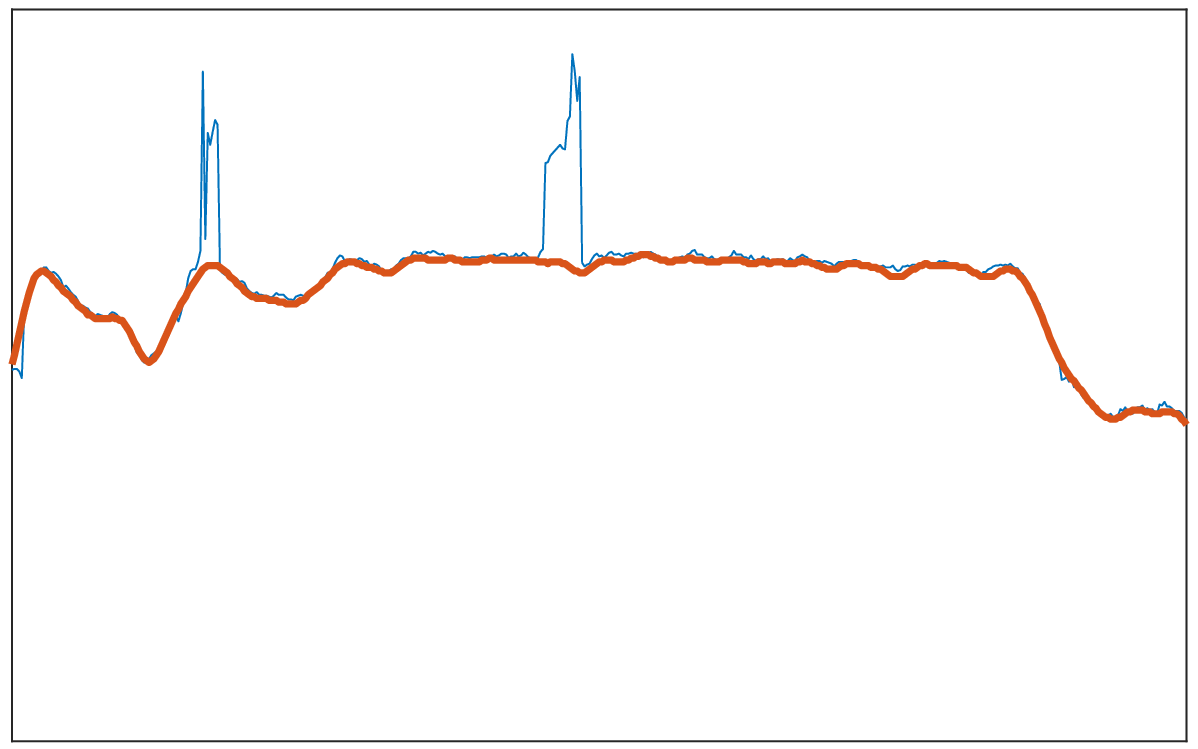}}
	\subfigure[$f_\text{WMLE}{[l]}$, $300\times 10^{-5}$.]{\includegraphics[width=1.39in]{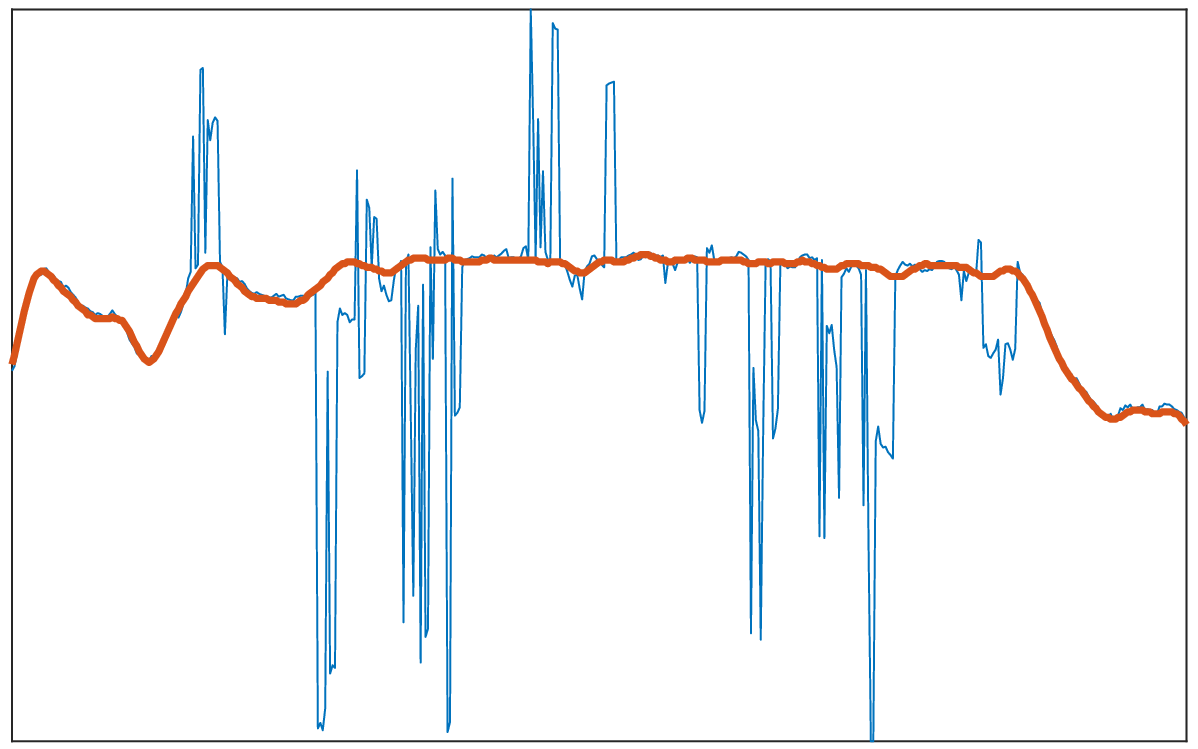}}
	\subfigure[$f_\text{E-MLE}{[l]}$, $11\times 10^{-5}$.]{\includegraphics[width=1.39in]{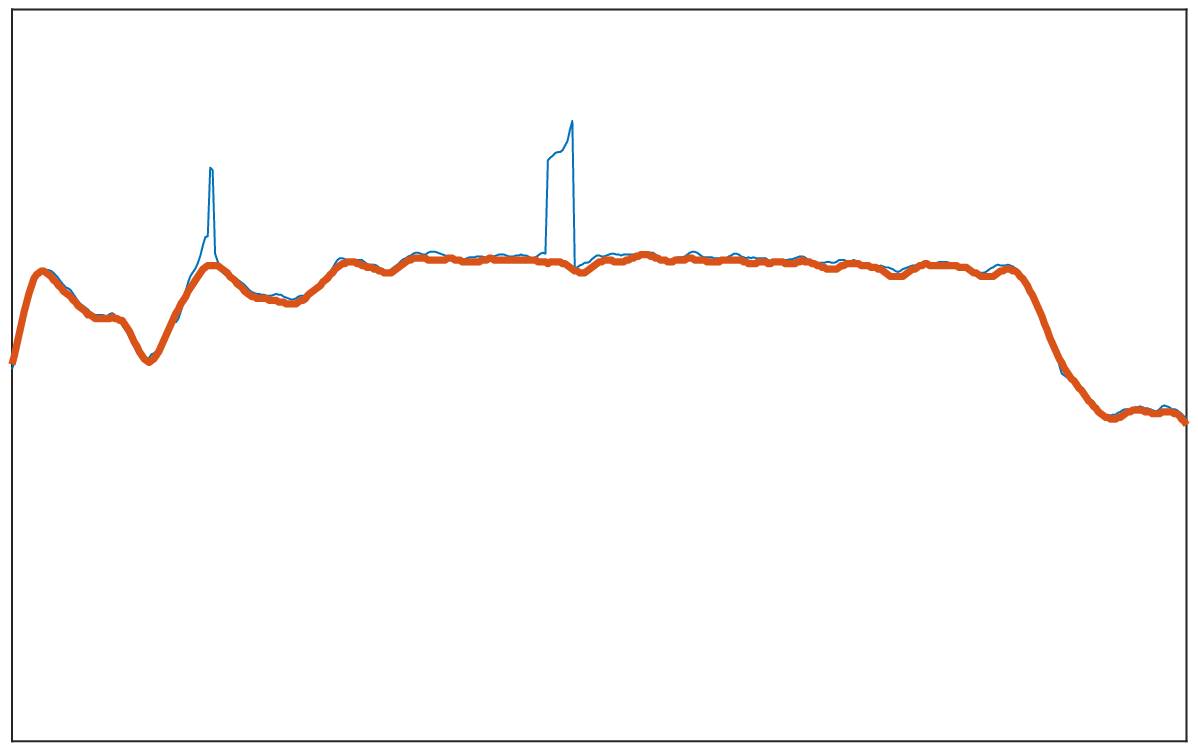}}\\
	\subfigure[$f_\text{E-WMLE}{[l]}$, $120\times 10^{-5}$.]{\includegraphics[width=1.39in]{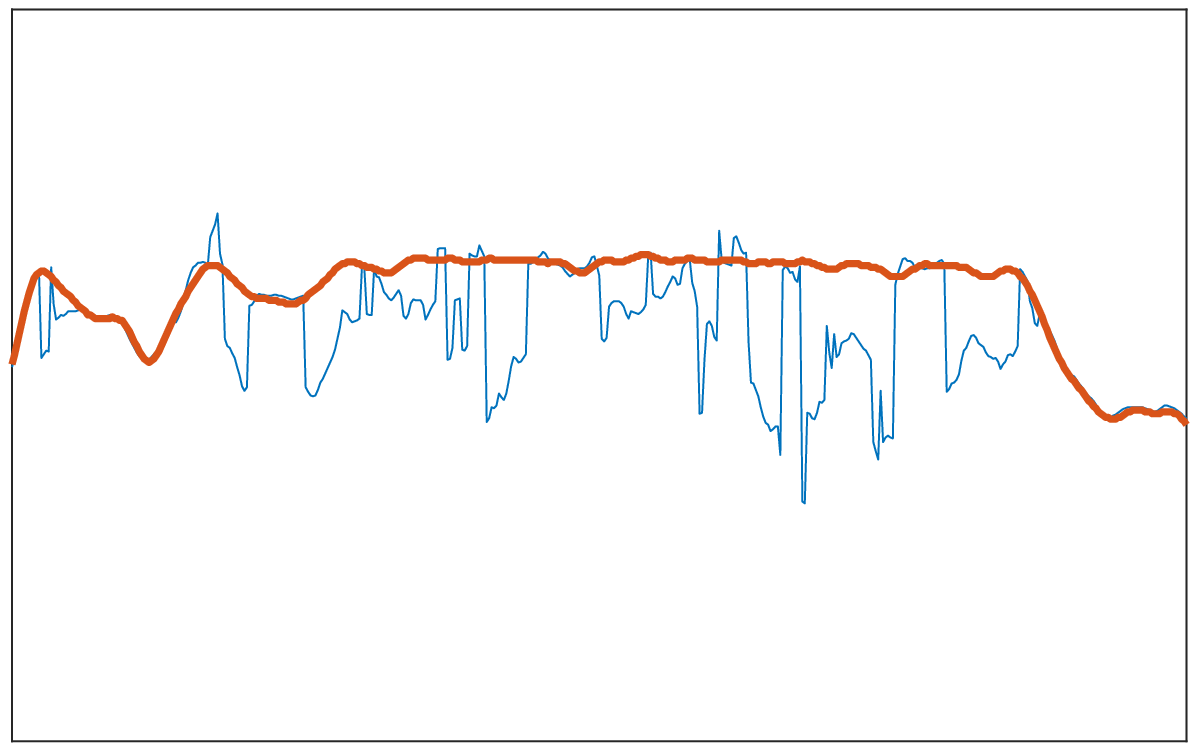}}
	\subfigure[$f_\text{S-MLE}{[l]}$, $0.72\times 10^{-5}$.]{\includegraphics[width=1.39in]{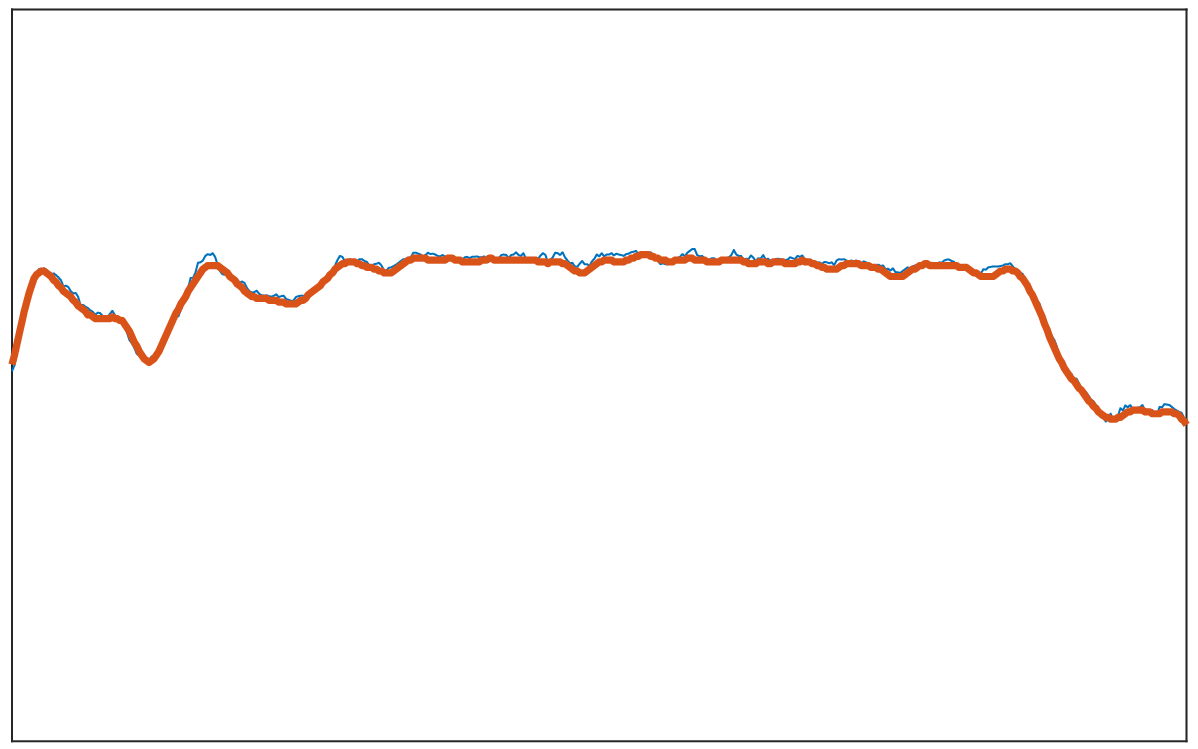}}
	\subfigure[$f_\text{S-WMLE}{[l]}$, $0.72\times 10^{-5}$.]{\includegraphics[width=1.39in]{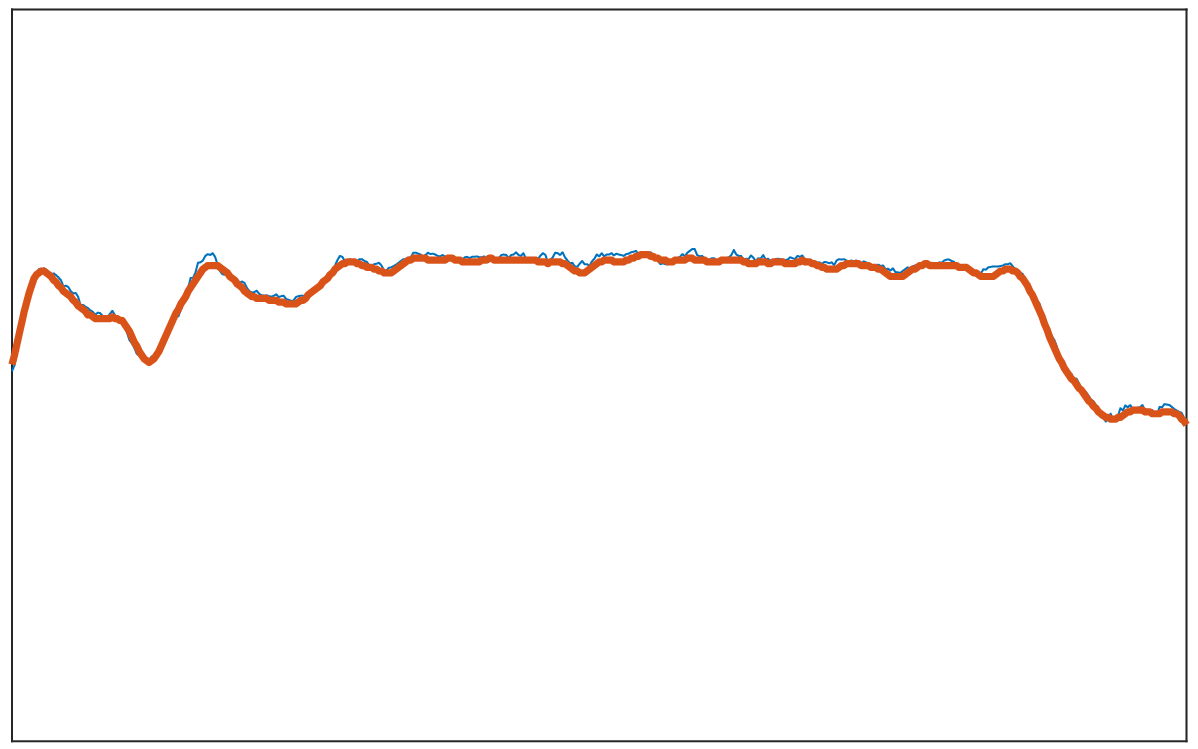}}
	\subfigure[$f_\text{P-MLE}{[l]}$, $0.68\times 10^{-5}$.]{\includegraphics[width=1.39in]{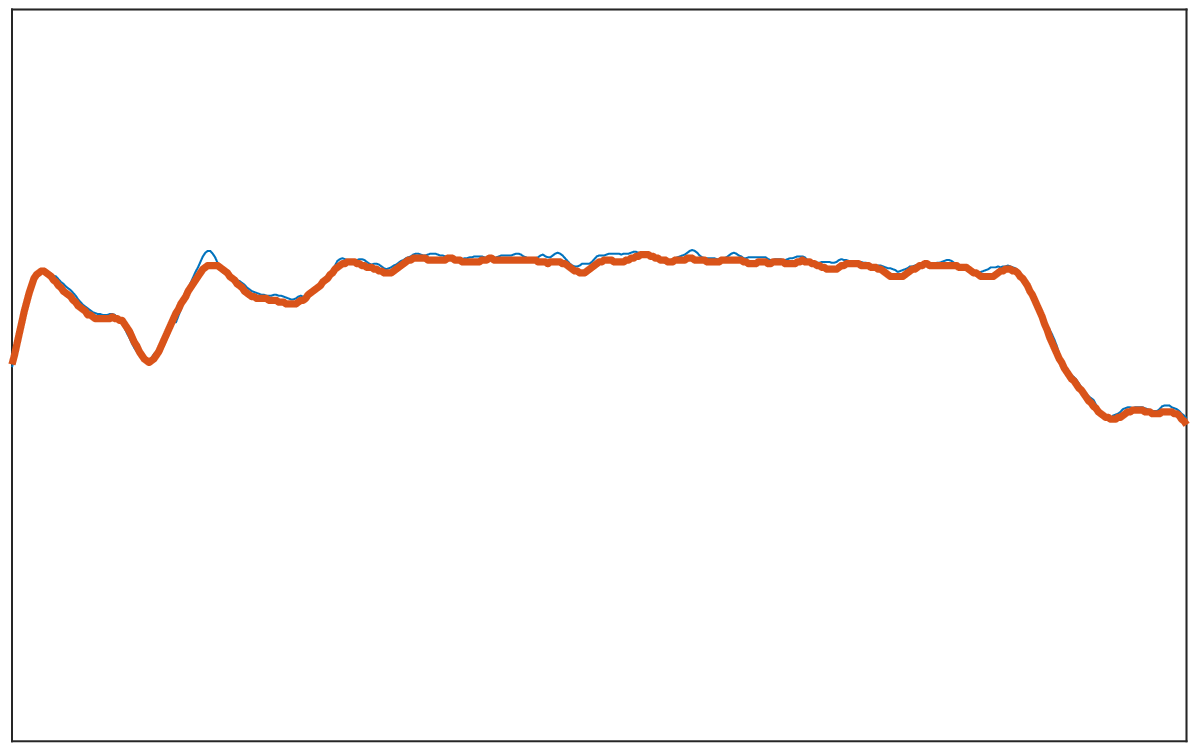}}
	\subfigure[$f_\text{P-WMLE}{[l]}$, $2.6\times 10^{-5}$.]{\includegraphics[width=1.39in]{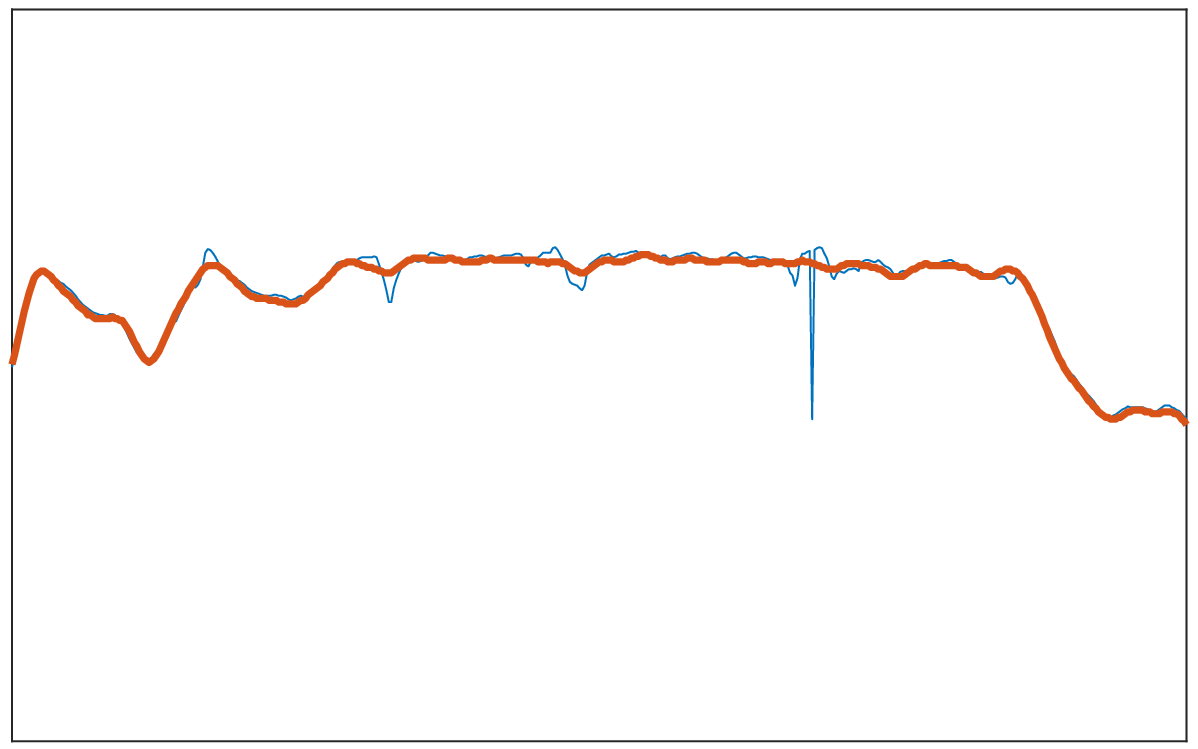}}
	\caption{A comparative demonstration of the single- and multi-tone ENF estimation schemes using an $8$-min real-world recording, $\kappa=4$, $\mathcal{M}=\{2\}$ for single-tone schemes, and $\mathcal{M}=\{2,3,4,5,6,7\}$ for multi-tone ones. All frequencies are normalized to the $2$nd harmonic band. For all subfigures, x-axis: sample index; y-axis: frequency (Hz); blue curve: estimated; red curve: ground truth. First row: individually evaluated harmonic components of the bandpass filtered real-world recording signal $x[n]$. Second row: individually evaluated harmonic components of $x_\text{E}[n]$, i.e., the signal after the HRFA. Last two rows: the ENF estimation results (sub-captions present scheme notation and the corresponding MSE values) of the $10$ schemes summarized in Table \ref{estimator_summary}. The indices of corrupted harmonic components are $\{2,3,5,6,7\}$. Before the HRFA, the GHSA yielded $\mathcal{C}=\{v_3\} \Rightarrow \Omega=\{4\}$ (first row gray one), while after the HRFA, the GHSA selected an extra enhanced component, $\mathcal{C}=\{v_1,v_3\}\Rightarrow \Omega=\{2,4\}$ (second row gray ones).}
	\label{Real_demo}
\end{figure*}

\subsection{Synthetic Analysis}
In this subsection, we provide a more comprehensive performance analysis based on Monte Carlo experiments using synthetic data. Specifically, WGN sequences are added to a synthetic ENF waveform to create noisy observations. Based on $100$ random realizations, the NMSE values versus SNR are presented in Fig. \ref{synthetic_MSE_SNR_ALL}, where marks are added to a few curves worth noting. The setting for the synthetic signals favors the MLE scheme because the noise is WGN with equal effects on each harmonic component, thus there is no harmonic component that is specifically corrupted. However, even under such a setting the MLE scheme, i.e., $f_\text{MLE}[l]$ \cite{Estimation_Harmonics} only yields the best performance when $\operatorname{SNR}\ge 5$ dB, while at very low SNR region, e.g., $\operatorname{SNR}\le -25$ dB, the scheme of $f_\text{E-WMLE}[l]$ achieves the lowest NMSE values. The advantage of the weighting mechanism of the WMLE scheme over the MLE scheme is observed only under very low SNR condition, while the HRFA is seen to be more effective in dealing with noise than the WMLE. The overall best performance is achieved by the proposed scheme plus the use of the MLE, i.e., $f_\text{P-MLE}[l]$, since it presents the lowest NMSE values in moderate SNR region and near the lowest values in high and low SNR regions. Besides, single-tone based methods yield inferior results against the multi-tone ones, where the former is almost the upper bound of the latter, and the single tone scheme with the RFA is strictly better than the one without it. The performances would become somewhat different in real-world situations when the noise is non-white and out of control, as will be illustrated in the next subsection.

Fig. \ref{synthetic_Omega_SNRs} further shows the numbers of selected harmonic indices before and after the use of  HRFA under different SNR conditions, thus illustrating the effectiveness of the enhancement module. Note that the noise in the synthetic signals is WGN, having statistically equal effects on the harmonic components, the HRFA thus tends to enhance either none or all components at the same time. 

\subsection{Experimental Results}
In this subsection, we evaluate the performances of the proposed ENF estimators using a real-world audio recording dataset that we have created around Wuhan University campus, termed as ENF-WHU dataset\footnote{The ENF-WHU Dataset and the Matlab programs of this paper could be downloaded from https://github.com/ghuawhu/ENF-WHU-Dataset.}. It contains $130$ audio recordings made in locations including classrooms, meeting rooms, graduate student offices, campus paths, main streets, dormitories, libraries, etc. The recording duration spans $5$ to $16$ minutes, and the recording environments include both sunny and rainy days and nights. All the recordings are resampled to $8000$ Hz with $16$-bit quantization and mono channel. The corresponding reference ENF segments (sampled at $400$ Hz) are extracted from the reference database and are also provided in the dataset.

Fig. \ref{Real_demo} provides a demonstrative example that compares the performances of the ENF estimation schemes summarized in Table \ref{estimator_summary} using a real-world recording from the ENF-WHU dataset. The results are summarized as follows. \textbf{i)} It can be seen from the first row that without the HRFA, only the $4$th harmonic component is selected for ENF estimation. Therefore, Fig. \ref{Real_demo} (c) is identical to Figs. \ref{Real_demo} (s) and (t). \textbf{ii)} Similar to Fig, \ref{synthetic_demo}, Figs. \ref{Real_demo} (a) and (g) are identical to Figs. \ref{Real_demo} (m) and (n) respectively. \textbf{iii)} From Figs. \ref{Real_demo} (o) and (p) we observe that directly applying multi-tone model based ENF estimation schemes \cite{Estimation_Harmonics,Estimation_Harmonics2}, without enhancement and harmonic selection, may increase the estimation error rather than reducing it. This is different from the synthetic example presented in Fig \ref{synthetic_demo} (o) in which the noise is WGN. \textbf{iv)} In this example it can be seen from the first row that $5$ harmonic components are severely corrupted, and the corresponding enhanced ones still look noisy from the second row. Therefore, after the HRFA, the GHSA only selected one additional component which is still noisy as shown in Fig. \ref{Real_demo} (g). Using the two components, $f_\text{P-MLE}[l]$ achieves the lowest MSE, indicating that the positive effect of adding the $2$nd component is more significant than the negative effect of the added $2$nd component being noisy. \textbf{v)} Generally, for such kind of very noisy recordings, a smart selection of harmonic components is more effective than signal enhancement only.

\begin{figure*}[!t]
	\centering
	\subfigure[$f_\text{P-MLE}{[l]}$ versus single-tone competitors.]{\includegraphics[width=3.4in]{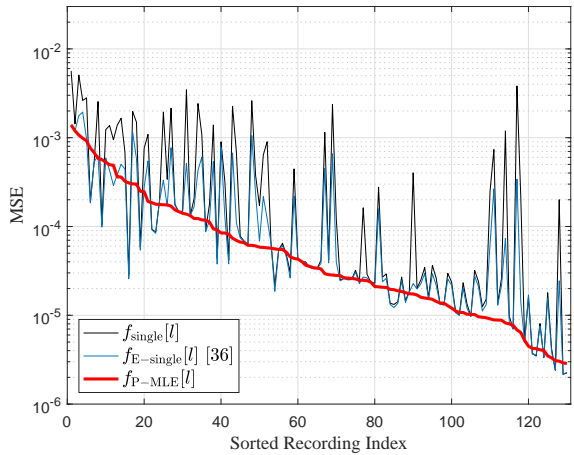}}
	\subfigure[$f_\text{P-MLE}{[l]}$ versus multi-tone competitors.]{\includegraphics[width=3.4in]{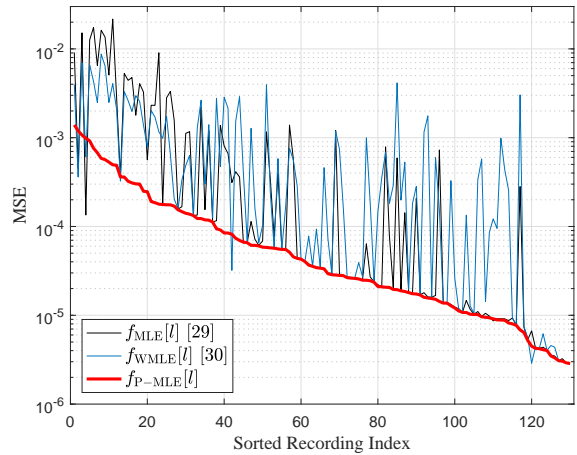}}
	\caption{Performance of the proposed $f_\text{P-MLE}[l]$ against signle- and multi-tone competitors using the ENF-WHU dataset, where $\mathcal{M}=\{2,3,4,5,6,7\}$ and $\kappa = 4$. Recording indices are arrange according to the descending order of the MSE values obtained by $f_\text{P-MLE}[l]$. For better visualization, results of the remaining schemes (see Table \ref{estimator_summary}) are not presented.}
	\label{Real_ALL_performance}
\end{figure*}

Experimental results using all the samples from the ENF-WHU dataset are provided in Fig. \ref{Real_ALL_performance}, where results of a few schemes with middle-ranked performances are not presented for better visualization. Fig. \ref{Real_ALL_performance} (a) and (b) present the comparative results of the proposed scheme against single- and multi-tone competitors respectively, where the red curves in both subfigures are identical to each other. The recording indices are adjusted according to the descending order of the MSE values obtained by $f_\text{P-MLE}[l]$. Results shown in these two subfigures reveal very important real-world situations when extracting the ENF from audio recordings. \textbf{i)} Quite different from our intuition and theoretical analysis (Figs. \ref{MSE_CRLB}, \ref{synthetic_demo}, and \ref{synthetic_MSE_SNR_ALL}), directly using multi-tone ENF estimators in fact resulted in worse results, which can be seen from the general higher errors in Fig. \ref{Real_ALL_performance} (b). This is because in practical situations, noise components at each harmonic band do not necessarily cancel each other, and using corrupted component will probably disturb rather than facilitate the estimation. \textbf{ii)} There exist a few cases in which the single-tone schemes are better than the proposed one. In such cases, components other than the $2$nd one are severely corrupted. After the HRFA, the enhanced components are accepted by the GHSA but the contained noise has more negative effects than the positive effect of including these components. To solve this problem, one may set a more stringent threshold for harmonic selection in (\ref{eta}). \textbf{iii)} In Fig. \ref{Real_ALL_performance} (b), the MLE and WMLE schemes directly use all the components without considering their usability. It can be seen that after enhancement and appropriately discarding unusable components, the MSE could be substantially reduced.

More quantitative results of all the ENF estimation schemes are provided in Table \ref{recording_performance_summary}, where $\overline{|\Omega|}$ is the averaged number of selected harmonic components based on the $130$ recordings in the ENF-WHU dataset, and $\operatorname{std}\{\cdot\}$ means standard deviation. In terms of harmonic selection, it can be seen that instead of the conventional settings that use either only the $2$nd or all $6$ components, the schemes with only the GHSA on average select $2.66$ components, while after the HRFA, the average number of selected components increases to $4.48$. In terms of estimation accuracy, the MLE scheme yielded the highest NMSE and most unstable MSE values, while the WMLE scheme slightly improves the overall performance, but the NMSE is still much higher than those of other competitors. Due to the fact that the $2$nd harmonic component is observed to be the strongest component in real-world audio recordings under Central China Grid, the single-tone method with the RFA \cite{Own_Singletone_Enhancement} has achieved very competitive results. The proposed scheme $f_\text{P-MLE}$ further reduces the NMSE achieved by $f_\text{E-single}[l]$ from $20\times 10^{-5}$ to $13\times 10^{-5}$ with the smallest standard deviation.

\begin{table}[!t]
	\renewcommand{\arraystretch}{1.3}
	\setlength{\tabcolsep}{6pt}
	\caption[]{Performances of ENF estimation schemes using real-world recordings from the ENF-WHU dataset, $\mathcal{M}=\{2\}$ for single-tone, $\mathcal{M}=\{2,3,4,5,6,7\}$ for multi-tone.}
	\label{recording_performance_summary}
	\centering
	\vspace*{-6pt}
	\begin{tabular}{l|c|c|l|l}
		\hline
		\hline
		\multicolumn{1}{c|}{{Scheme}}  & $|\mathcal{M}|$ & $\overline{|\Omega|}$ & \multicolumn{1}{c|}{NMSE} & \multicolumn{1}{c}{$\operatorname{std}\{\text{MSE}\}$}\\
		\hline
		$f_\text{single}[l]$ & $1$ & $1$ & $54\times 10^{-5}$ & $10\times10^{-4}$ \\
		$f_\text{E-single}[l]$ \cite{Own_Singletone_Enhancement} & $1$ & $1$ & $20\times 10^{-5}$ & $3.4\times 10^{-4}$\\
		$f_\text{MLE}[l]$ \cite{Estimation_Harmonics}  & $6$ & $6$ & $140\times 10^{-5}$ & $37\times10^{-4}$ \\	
		$f_\text{WMLE}[l]$ \cite{Estimation_Harmonics2} & $6$ & $6$ & $95\times 10^{-5}$ & $16\times10^{-4}$\\
		\hline
		$f_\text{E-MLE}[l]$ & $6$ & $6$ & $15\times 10^{-5}$ & $2.9\times 10^{-4}$\\
		$f_\text{E-WMLE}[l]$ & $6$ & $6$ & $34\times 10^{-5}$ & $3.2\times 10^{-4}$ \\
		$f_\text{S-MLE}[l]$ & $6$ & $2.66$ & $34\times10^{-5}$ & $8.1\times10^{-4}$\\
		$f_\text{S-WMLE}[l]$ & $6$ & $2.66$ & $28\times10^{-5}$ & $7.0\times10^{-4}$ \\
		\hline
		$f_\text{P-MLE}[l]$ & $6$ & $\mathbf{4.48}$ & $\mathbf{13\times 10^{-5}}$& $\mathbf{2.4\times 10^{-4}}$\\
		$f_\text{P-WMLE}[l]$ & $6$ & $4.48$ & $23\times 10^{-5}$ & $2.5\times 10^{-4}$\\
		\hline
		\hline
	\end{tabular}
\end{table}

\section{Conclusion}
In this paper, we have presented a framework for robust ENF extraction from real-world audio recordings, consisting of two proposed algorithms, i.e., the HRFA and GHSA, for ENF signal enhancement and harmonic selection respectively. The HRFA is an extension of the RFA previously proposed for single-tone ENF enhancement, which is shown to be free of cross-component interference thanks to the spectral property of the ENF. The harmonic selection problem is formulated as an MWC problem in graph theory which is then effectively solved with the use of the BKA. Combined with the state-of-the-art MLE for ENF estimation, the proposed scheme, denoted by $f_\text{P-MLE}[l]$, yields the best performance against other single- and multi- tone based competitors. Using our ENF-WHU dataset containing $130$ real-world audio recordings, the advantages of thhe proposed scheme have been extensively verified. Some important results are summarized as follows.
\begin{itemize}
	\item In real-world situation, using more harmonic components for ENF estimation does not guarantee the theoretical performance gain, because of the non-stationary noise.
	\item When adding one more harmonic component, one needs to evaluate whether the performance gain of adding this component is more significant than the interference caused by the noise in it.
	\item On the one hand, the proposed framework utilizes the HRFA to preserve more harmonic components for theoretical performance gain. On the other hand, via the GHSA, the framework discards the components that may bring more interference than theoretical gain. Therefore, it achieves the most robust ENF estimation performance.  
\end{itemize}
Future research efforts could be devoted to the replacement of (\ref{eta}) with more effective thresholding mechanisms. Note that currently $\eta$ is only related with the test recording by the duration $N_\text{ENF}$. In fact, more information about the test recording could be incorporated to determine $\eta$. For example, considering the noise condition, one could tend to select less harmonic components (setting a higher $\eta$) if they are observed to be very noisy and to select more components (setting a lower $\eta$) if they are observed to be clean. Following this direction, it is possible to establish an optimal harmonic selection mechanism such that the corresponding estimation error could always be the minimized compared to the competitors.

\ifCLASSOPTIONcaptionsoff
\newpage
\fi

\bibliographystyle{IEEEtran}
\bibliography{ENFrefs}

\end{document}